\documentclass[conference]{IEEEtran}
\IEEEoverridecommandlockouts
\usepackage{cite}
\usepackage{amsmath,amssymb,amsfonts}
\usepackage{algorithmic}
\usepackage{graphicx}
\usepackage{textcomp}
\usepackage{xcolor}
\def\BibTeX{{\rm B\kern-.05em{\sc i\kern-.025em b}\kern-.08em
    T\kern-.1667em\lower.7ex\hbox{E}\kern-.125emX}}

\usepackage{hyperref, xcolor}
\usepackage{multirow}
\usepackage{multicol}
\usepackage{xspace}
\usepackage{wrapfig}

\usepackage{enumitem}
\usepackage{wrapfig,booktabs}
\usepackage{float}
\usepackage{algorithm}

\usepackage{mathtools}
\usepackage[caption=false]{subfig}

\usepackage{lipsum} 
\usepackage[T1]{fontenc}

\newcommand{\globalmodel}{global model}
\newcommand{\modelowner} {model owner}

\newcommand{\client}{client}
\newcommand{\aggregator}{aggregator}
\newcommand{\ourmethod}{\textsc{Waffle}\xspace}

\newcommand{\ourpattern}{\textsc{WafflePattern}\xspace}
\newcommand{\agg}{$\mathit{Agg}$}
\newcommand{\aggt}{$\mathit{Agg}$\text{ }}

\newcommand{\newtext}[1]{{#1}}
\definecolor{greena}{rgb}{0.0, 0.7, 0.1}
\newcommand{\removetext}[1]{}

\begin{document}

\title{\ourmethod: Watermarking in Federated Learning\\
\thanks{\newtext{This work was supported in part by Intel (in the context of the Private-AI Institute) and SAPPAN, a project funded by the European Union's Horizon 2020 Research and Innovation Programme under Grant Agreement no. 833418. We also thank Sebastian Szyller for interesting discussions, and Aalto Science-IT project for computational resources.}}}

\author{
	\IEEEauthorblockN{Buse G. A. Tekgul\IEEEauthorrefmark{1}, 
		Yuxi Xia\IEEEauthorrefmark{1}, 
		Samuel Marchal\IEEEauthorrefmark{2}, 
		and N. Asokan \IEEEauthorrefmark{3}\IEEEauthorrefmark{1}}
	\IEEEauthorblockA{\IEEEauthorrefmark{1}Department of Computer Science, Aalto University, 02150, Espoo, Finland\\Email: {buse.atlitekgul@aalto.fi, yuxi.xia.work@gmail.com}}
	\IEEEauthorblockA{\IEEEauthorrefmark{2}F-Secure Corporation, 00180, Helsinki, Finland \\Email: samuel.marchal@aalto.fi}
	\IEEEauthorblockA{\IEEEauthorrefmark{3} University of Waterloo, Waterloo, ON N2L 3G1, Canada \\Email: asokan@acm.org}}	

\maketitle

\begin{abstract}
\emph{Federated learning} is a distributed learning technique where machine learning models are trained on client devices in which the local training data resides. The training is coordinated via a central server which is, typically, controlled by the intended owner of the resulting model. By avoiding the need to transport the training data to the central server, federated learning improves privacy and efficiency. But it raises the risk of \emph{model theft} by clients because the resulting model is available on every client device. Even if the application software used for local training may attempt to prevent direct access to the model, a malicious client may bypass any such restrictions by reverse engineering the application software. \emph{Watermarking} is a well-known deterrence method against model theft by providing the means for model owners to demonstrate ownership of their models. Several recent deep neural network (DNN) watermarking techniques use \emph{backdooring}: training the models with additional mislabeled data. Backdooring requires full access to the training data and control of the training process. This is feasible when a single party trains the model in a centralized manner, but not in a federated learning setting where the training process and training data are distributed among several client devices. In this paper, we present \ourmethod, the first approach to watermark DNN models trained using federated learning. It introduces a re-training step at the server after each aggregation of local models into the global model. We show that \ourmethod\ efficiently embeds a resilient watermark into models incurring only negligible degradation in test accuracy ($-0.17\%$), and does not require access to  training data. We also introduce a novel technique to generate the backdoor used as a watermark. It outperforms prior techniques, imposing no communication, and low computational ($+3.2\%$) overhead\footnote{The code for reproducing our work can be found at \url{https://github.com/ssg-research/WAFFLE}}.
\end{abstract}


\section{Introduction}\label{sec:intro}
Distributed machine learning has gained considerable attention in many big data processing applications such as recommendation systems, smart assistants, and health-care, due to its efficient parallelization and scalability properties. Federated learning~\cite{mcmahan2016communication} is an instance of privacy-preserving distributed machine learning that allows decentralized training of deep neural networks (DNNs) by many parties holding local data samples. Major companies have already utilized federated learning in various large-scale applications such as improving the Google Keyboard (Gboard) suggestions~\cite{hard2018federated,yang2018applied} or Apple's voice recognition models~\cite{improving-on-device-speaker}. In these applications, clients only need to download a set of tools or a specific software (e.g., Tensorflow Lite\footnote{ML for Mobile and Edge Devices, https://www.tensorflow.org/lite/}) into their mobile devices, and execute the training with the on-device, cached data. Hence, instead of using publicly available or synthetic datasets that often do not represent the real-world distribution, model owners can deploy high-quality models with no access to the possibly sensitive training data.

The most common federated learning setting in commercial, global-scale applications is \textit{client-server} involving three parties: \textit{\modelowner}, \textit{\aggregator} (hosted on a server), and a large number of \textit{\client s} training a common \textit{\globalmodel}~\cite{bonawitz2019towards}. 
Federated learning consists of iterative \textit{aggregation rounds} where (1) the \aggregator\ sends the \globalmodel\ parameters to \client s, (2) each \client\ replaces parameters of its local model using the \globalmodel, re-trains it using the on-device data, and sends updated parameters back to the \aggregator, which (3) combines them into a new \globalmodel. The final \globalmodel\ is delivered to the \modelowner\ when the training is completed. Since each client's training data stays in place, federated learning preserves privacy of \client s' data and avoids incurring substantial costs for transferring training data from \client s to a centralized trainer. 

Despite its advantages, federated learning brings forth the issue of preserving ownership. In large-scale, client-server federated learning applications, there are multiple clients who are data-owners and \emph{can use the model} in their local devices, but there is \emph{only one \modelowner.}
However, a side-effect of the training process is that each client can gain complete access to the \globalmodel\ in every round, including the final one by reverse engineering the application software~\cite{li2018reverse} or on-device dynamic analysis~\cite{sun2020mind}. Although the model owner can deploy further mechanisms for hiding model parameters during the local training (e.g. homomorphic encryption), this might lead to high latency, bandwidth costs, and additional requirements for benign clients~\cite{kairouz2019advances}. It is therefore useful to have a means of demonstrating the intended ownership in case a malicious client uses the \globalmodel\ residing on their devices in unauthorized ways such as monetizing the model or making it available to their own customers.

Recently, different watermarking techniques~\cite{adi2018turning,guo2018watermarking,szyller2019dawn,uchida2017embeeding,zhang2018protecting}  have been proposed to demonstrate ownership of DNN models. In a typical backdoor-based watermarking procedure, the \modelowner\ first designs a secret \emph{watermark} which consists of mislabeled input output pairs. Then, the \modelowner\ trains the model with both the training dataset and the watermark in order to embed the watermark into the model. This watermark can be subsequently used to demonstrate ownership.  
However, existing DNN watermarking solutions cannot be directly applied in federated learning for two reasons. 
First, model owners cannot use techniques for generating watermarks using training data because they lack access to the training data (e.g., Gboard~\cite{yang2018applied}). Second, training is performed in parallel by several \client s, some of them being potentially malicious. \newtext{Although some ``trusted clients'' can join the watermark embedding phase, the training setup and aggregation rules have to be changed to handle them differently.  Thus, watermark embedding cannot be implemented easily by clients.}
Our goal is to design a procedure for effectively embedding watermarks into DNN models trained via client-server federated learning, without decreasing the accuracy of the resulting \globalmodel\ while minimizing the computational and communication overhead imposed on the distributed training process. 
We claim the following contributions:

\begin{enumerate}[labelindent=0pt]
\item \textbf{Problem definition}: We identify the problem of demonstrating ownership of models trained via client-server federated learning and define requirements for a solution addressing this problem (Section~\ref{ssec:requirements}).
\item \textbf{Watermarking procedure for federated learning}: We introduce \ourmethod: the first solution for addressing ownership problem in client-server federated learning (Section~\ref{sec:method}). \ourmethod\ leverages capabilities of the \aggregator\ to embed a backdoor-based watermark~\cite{adi2018turning} by re-training the \globalmodel\ with the watermark during each aggregation round. We show on MNIST and CIFAR10 using two DNN architectures that \ourmethod\ embeds watermarks without degrading the model performance (Section~\ref{ssec:modelutility}).
\item \textbf{Data-independent watermark generation method}: We introduce \ourpattern, a novel data-independent method to generate watermarks for DNN image classification models (Section~\ref{ssec:watermarkgeneration}). It generates images having random but class-consistent patterns on a random background, which is suitable for federated learning.
Compared to prior watermark generation methods~\cite{adi2018turning,zhang2018protecting,rouhani2018deepsigns}, \ourpattern retains model performance better (Section~\ref{ssec:modelutility}) and imposes lower communication and computational overhead (Section~\ref{ssec:evaloverhead}).
We also show that \ourmethod with \ourpattern is resilient to watermark removal techniques including fine-tuning, pruning and reverse-engineering (Section~\ref{ssec:wmrobustness}) if no more than 10\% of the clients collude to defeat the watermarking procedure. 
\end{enumerate}

Our procedure focuses on embedding effective watermarks into models trained with federated learning. We do not discuss the demonstration of ownership using our watermarks because secure schemes~\cite{adi2018turning,szyller2019dawn} for registering and using backdoor-based watermarks  to reliably demonstrate ownership of DNN models already exist and these schemes can be used with our watermarks. 
We also leave discussions on the legal validity of DNN model watermarks as out of scope.

\section{Background}\label{background}

\removetext{\subsection{Deep Neural Network}
A DNN is a hierarchical composition of neural network layers. It is a parameterized function 
 $F(x, w): \mathbb{R}^n \rightarrow  \mathbb{R}^m$ that maps an input $x \in \mathbb{R}^n$ to an output vector $y \in \mathbb{R}^m$, where $n$ refers to the number of input features and $m$ is the length of the output vector. $w$ is the parameter vector learned by minimizing the training loss over a labeled training dataset. In classification problems, the predicted class $y$ is obtained by $y =argmax (F(x,w))$. DNNs are natural function approximators and their performance can be measured with the test accuracy $Acc(F, D_{test})$, which gives the percentage of the correctly classified samples on a labeled test dataset $D_{test}=\{x_i, y_i\}_{i=1}^N$. For simplicity, we will use $w$ to denote both model parameters and the model itself. 
 }

\subsection{Federated Learning}\label{ssec:FederatedLearning}
The client-server federated learning is composed of three main parties: 1) A large number of \client s $C = \{c_j\}_{j=1}^K$ who are data owners and keep their datasets $D_{c_j}$ private, 2) a \modelowner\ $O$ providing a randomly initialized global model $w_G$ at the beginning of federated learning and getting it trained at the end, and 3) a secure aggregator \agg~\cite{bonawitz2017practical} located between $O$ and $C$. We focus on training a DNN model, which is a function $F(x, w): \mathbb{R}^n \rightarrow  \mathbb{R}^m$ that we simply refer to as its parameters $w$, e.g., $w_G$.
$O$ cannot obtain any information about $D_{c_j}$ due to the secure aggregation protocol~\cite{bonawitz2017practical} implemented by \agg. In this work, we focus on federated learning using the FederatedAveraging (FedAvg) algorithm~\cite{mcmahan2016communication}, a widely used aggregation rule. 
In FedAvg, \client s train their models using stochastic gradient descent (SGD). 
Before federated learning starts, $w_G$ is initialized by $O$. In aggregation round $t$,
\begin{enumerate}
\item \aggt sends $w_{G(t)}$ to a subset of clients $C_{sub} = \{c_i\}_{i=1}^L$, where  
$(L \ll K)$. 
\item Each $c_i$ updates $w_{c_i(t-1)}$ with $w_{G(t)}$, re-trains the updated $w_{c_i(t)}$ by a pre-determined number of local passes over its local dataset $D_{c_{i}}$, and sends the re-trained local model $w_{c_i(t)}$ to \agg.
\item \aggt averages all local models into a new $w_{G(t+1)}$. 
\end{enumerate}

\subsection{Watermarking DNN Models by Backdooring}\label{ssec:watermarkingDNNmodels}
Several works~\cite{adi2018turning,rouhani2018deepsigns,guo2018watermarking,li2019prove,uchida2017embeeding,zhang2018protecting} have proven the feasibility of embedding watermarks into DNNs for demonstration of ownership. In general, a watermark set $\mathrm{WM}_w$ consists of samples $\{x, B(x)\}$ designed by the owner of the DNN model $w$. The model owner embeds $\mathrm{WM}_w$ into $w$ 
by optimizing $w$ on both the training set and $\mathrm{WM}_w$ such that $B(x) = w^{+}(x)$ and $B(x) \neq w(x)$ for almost all $x \in\mathrm{WM}_w$, where $w^{+}(x)$ is the watermarked model.
If the \modelowner\ suspects that another model $w_{adv}$ is possibly derived from $w^{+}$, $\mathrm{WM}_w$ with a pre-defined verification algorithm $\textsc{Verify}$ is used to demonstrate ownership if $\textsc{Verify}(w_{adv}, \mathrm{WM}_{w})$ returns $True$.

In this paper, we focus on watermarking via 
backdooring~\cite{adi2018turning,guo2018watermarking,zhang2018protecting}. 
A \textit{backdoor}~\cite{liu2018trojaning} consists of a \textit{trigger set} of samples with incorrect labels. 
In computer vision, a trigger set usually contains specific patterns that can be added to an image~\cite{guo2018watermarking} or a set of images unrelated to the actual task~\cite{adi2018turning,guo2018watermarking}, both of which serve a similar purpose. The backdoor is injected into a DNN by training it with both the correct training data and the backdoor. At inference time, the backdoored DNN performs normally on clean inputs but returns the expected incorrect label when a sample from the trigger set is given as an input.
To reliably demonstrate that a model $w_{adv}$ is derived from $w^{+}$, i.e., to obtain $\textsc{Verify}(w_{adv}, \mathrm{WM}_{w}) \rightarrow True$, $\mathrm{WM}_{w}$ must be secret and registered in a timestamped public bulletin~\cite{adi2018turning,szyller2019dawn}. In addition, the accuracy of $w_{adv}$ on $\mathrm{WM}_{w}$ must be over a threshold $T_{acc}$:  $Acc(w_{adv}, \mathrm{WM}_{w}) \geq T_{acc}$. The value for $T_{acc}$ is computed using the size of the watermark set $|\mathrm{WM}_{w}|$ and the number of classes $m$ for $w$~\cite{adi2018turning,szyller2019dawn}.
\section{Ownership Demonstration in Federated Learning}\label{sec:problem}

\subsection{Adversary Model} \label{ssec:adversarymodel}
\begin{figure}[t]
 \centering
  \includegraphics[width=1.0\columnwidth]{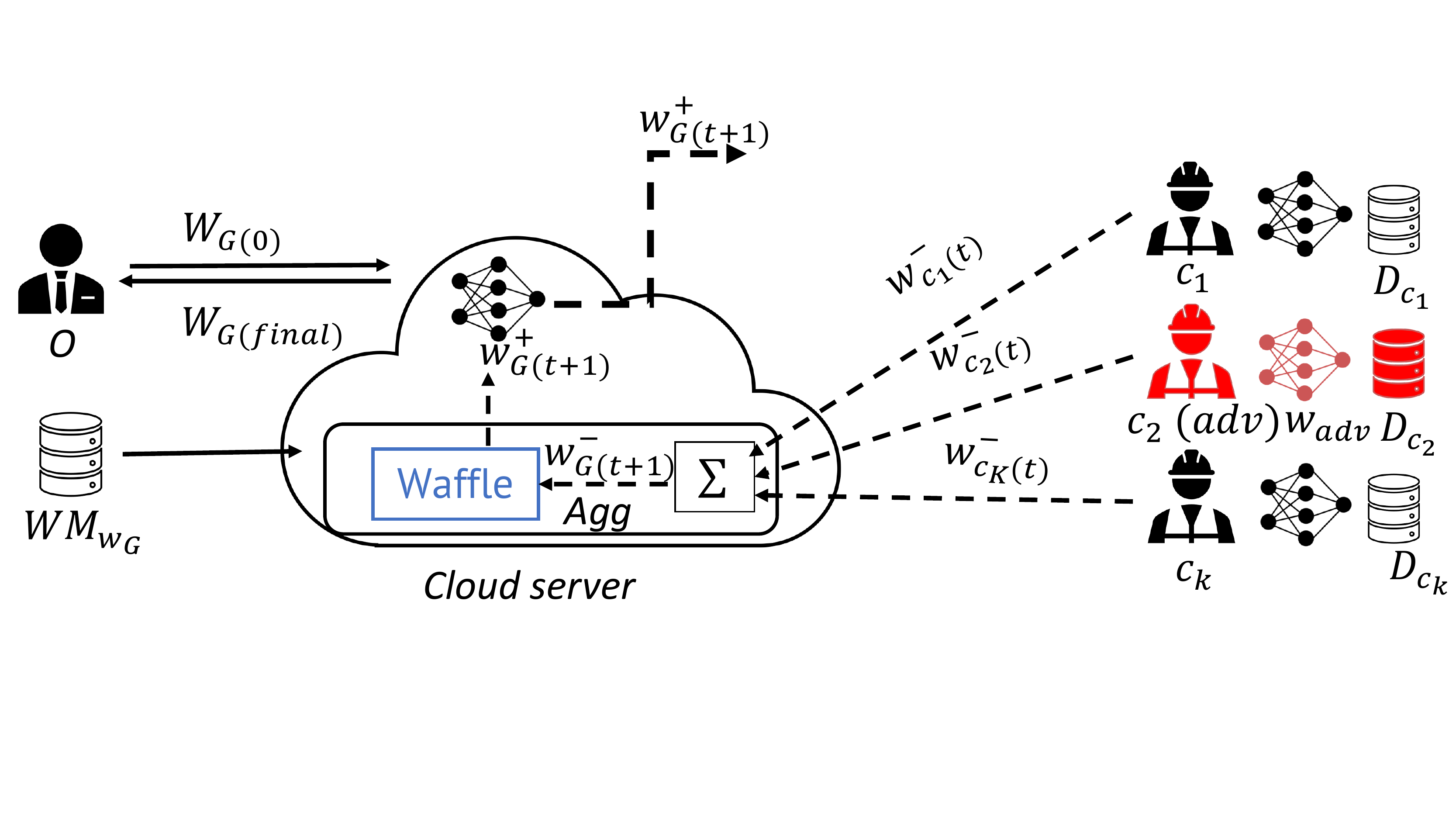}
  \vspace{-40pt}
  \caption{Architecture and adversary model for watermarking in federated learning. Parties under the control of an adversary are highlighted in red.}\label{fig:advmodel}
\end{figure}

The ability to demonstrate ownership of machine learning models has emerged as an important concern because models can be monetized and represent a business advantage~\cite{juuti2019prada,szyller2019dawn,tramer2016stealing}.
Figure~\ref{fig:advmodel} shows our adversary model. 
We consider malicious \client s to be our primary adversaries. 
Following prior work~\cite{bonawitz2017practical}, we assume that \agg\ is the only trusted party in federated learning. \agg\ resides in a central server, and is defined as an incorruptible third party~\cite{bonawitz2017practical,fl} between $C$ and $O$. In some applications, the cloud server can belong to $O$ (e.g., 
Gboard~\cite{yang2018applied}, Apple's Siri~\cite{improving-on-device-speaker}) but \agg\ is still considered a \textit{secure} party: it implements a secure aggregation protocol~\cite{bonawitz2017practical} and uses cryptographic primitives while routing messages between the server and clients.
In unauthenticated or unencrypted network models like federated learning, \agg\  provides the strongest possible security for protecting data confidentiality~\cite{bonawitz2019towards,bonawitz2017practical,fl}.
\subsubsection{Adversaries' goal} Adversaries aim to obtain a $w_{adv}$ having the same performance as $w_G$ on a test dataset $D_{test}$, so that
$Acc(w_{adv}, D_{test}) \approx Acc(w_{G}, D_{test})$ while evading watermark detection, i.e., 
$\textsc{Verify}(w_{adv}, \mathrm{WM}_{w_{G}})\rightarrow False$, where $\mathrm{WM}_{w_{G}}$ is the watermark generated by $O$.
We assume that an adversary $adv = c_i$ has the following capabilities:
\begin{itemize}
 \item has access to its own relevant training data $D_{adv}$, but does not have access to other \client s' training data. Each $D_{c_i}$ is independent, identically distributed (IID), balanced and limited in size such that every $c_i$ is incentivized to participate in federated learning. $adv$ having a large $D_{adv}$ would rather train its own model instead of stealing $w_G$. 
 \item gains white-box access to $w_{adv(t)}$, and can save $w_{adv(t)}$ at any aggregation round $t$.
 \item can apply any post processing technique (e.g., fine-tuning, pruning, reverse engineering) to 
$w_{adv(t)}$  in order to remove watermarks. 
\end{itemize}

\subsubsection{Assumptions} We assume that every $c_i$ is incentivized to maximize the accuracy of $w_G$. $c_i$ does not backdoor~\cite{bagdasaryan2018backdoor}, poison or embed its own watermark sets in $w_G$. These attacks might decrease the performance of $w_G$ and $w_{adv}$, which is contrary to the goal of every $c_i$, including $adv$, who aims to obtain a model $w_G / w_{adv}$ having the best performance possible. Furthermore, \client s does not change the training scheme and use the same hyper-parameters provided by $O$ (e.g., number of local passes $E_c$, loss function $\ell_c$, learning rate $\eta_c$, batch size $|b_c|$), since they are optimized to maximize the accuracy of $w_G$.

\subsection{Requirements}\label{ssec:requirements}
We define the following requirements for designing an effective watermarking scheme (\textbf{W1:3}) as well as performance conditions (\textbf{P1:3}) which must be met in federated learning. Performance conditions should be also taken into account for building other type of solutions for ownership demonstration of federated learning models. 

\begin{enumerate}[label=\textbf{W\arabic*}, nolistsep]
 \item\label{req:w_ownership} \textbf{Demonstration of ownership:}At any aggregation round $t$, $\textsc{Verify}(w_{adv(t)}, \mathrm{WM}_{w_{G}})$ should return $True$ , which requires $Acc(w_{adv(t)}, \mathrm{WM}_{w_{G}}) \geq T_{acc}$. 
 \item \label{req:w_robustness} \textbf{Robustness:} $\mathrm{WM}_{w_{G}}$ embedded in $w_{adv}$ should be resilient against attacks that try to remove it without destroying the performance of $w_{adv}$. This requirement is satisfied when~\ref{req:w_ownership} still holds or $Acc(w_{adv}, D_{test}) \ll Acc(w^{+}_{adv}, D_{test})$ after applying a watermark removal attack to $w_{adv}$.
\item \label{req:w_data_independence} \textbf{Data independence:} Watermarking procedure should not require training data knowledge, and $O$ cannot use clients' training data in order to generate their own watermark set $\mathrm{WM}_{w_{G}}$. 
\end{enumerate}
\begin{enumerate}[label=\textbf{P\arabic*}]
	\item \label{req:p_utility} \textbf{Model utility:} Watermarking should not degrade the performance of the local model, i.e., 
$Acc(w^{+}_{adv}, D_{test}) \approx Acc(w_{adv}, D_{test})$.  
	\item \label{req:p_communication} \textbf{Low communication overhead:} Embedding $\mathrm{WM}_{w_{G}}$ should not increase the total number of aggregation rounds or the amount of the data exchanged to reach the convergence of $w_G$. 
   \item \label{req:p_computation} \textbf{Low computation overhead:} Implementing watermark embedding with $\mathrm{WM}_{w_{G}}$ inside any party should incur minimal additional computation. 
\end{enumerate}
\section{Watermarking in Federated Learning}\label{sec:method}
\subsection{Current Challenges}\label{ssec:wmoverview}
Watermarking can help \modelowner s to prove ownership of their DNN models.
However, state-of-the-art techniques~\cite{adi2018turning,guo2018watermarking,uchida2017embeeding,zhang2018protecting} cannot be directly implemented in client-server federated learning, since they assume full control over the training process as in centralized machine learning. Moreover, \client s 
might be malicious; therefore, they should be excluded from any involvement in watermarking and should not have access to the watermark set. 
$O$ has two options for embedding the watermark using existing techniques:
\begin{itemize}
\item  ``Pre-embedding'' into $w_{G(0)}$ before training starts.
\item ``Post-embedding'' into $w_{G(t)}$ before deploying the model.
\end{itemize}
Pre-embedding achieves a watermark accuracy of $100\%$ during the first aggregation round, while Post-embedding achieves that at the last aggregation round $t$. However, both techniques have limitations. Pre-embedded watermarks are easily removed from the global model after several aggregation rounds. Post-embedding demonstrates a reliable proof of ownership when the model is deployed, but $adv$ can use updates from the global model, $w_{G(t-1)}$, one aggregation round before the training is completed. $w_{adv(t-1)} = w_{G(t-1)}$ contains no watermark and $Acc(w_{adv(t-1)}, D_{test}) \approx Acc(w_{adv(t)}, D_{test})$. Finally, post-embedded watermarks are not resilient to removal attacks such as fine-tuning and pruning
as shown in~\cite{adi2018turning}. 
\textbf{Both techniques fail to satisfy~\ref{req:w_ownership} and~\ref{req:w_robustness} and are not feasible for watermarking federated learning models.}
\subsection{\ourmethod\ Procedure}\label{ssec:proposedapp}
Considering the capabilities of \agg\ presented in Section~\ref{ssec:adversarymodel}, we add a new responsibility for it: \emph{watermarking}. In \ourmethod, $O$ shares its watermark set $\mathrm{WM}_{w_G}$ only with \agg. \aggt re-trains $w_{G}$ to guarantee that the watermark is embedded and $\textsc{Verify}$ returns $True$ for watermarked models: $w^{+}_{G(t)}$ and $w^{+}_{c_{i}(t)}$. 
\ourmethod\ makes \textbf{no modification to client operations or the secure aggregation.} Inside \agg, we introduce two new functions: \textsc{Pretrain} and \textsc{Retrain}. \textsc{Pretrain} is a one time operation performed before federated learning starts and \textsc{Retrain} is a recurrent operation performed at each aggregation round $t$. The pseudo-code for \textsc{Pretrain} and \textsc{Retrain} is given in Algorithm~\ref{alg:algorithmPR} (Appendix~\ref{ssec:algorithmWaffle}). 

\textsc{Pretrain} gets a randomly initialized $w_{G(0)}$ and returns a watermarked version $w^+_{G(0)}$, $Acc(w^+_G(0), \mathrm{WM}_{w_G}) = 100\%$. \textsc{Retrain} first implements the secure aggregation: averaging re-trained  $w^{-}_{c_i(t)}$\footnote{$w^{-}$ refers to a re-trained version of a previously watermarked model $w^{+}$.} received from $C_{sub}$ and updating $w^-_{G(t+1)}$ with this average. Then, it re-trains $w^-_{G(t+1)}$ using $\mathrm{WM}_{w_G}$ until $Acc(w^-_{G(t+1)}, \mathrm{WM}_{w_G})$ reaches threshold value $th$. We also define the maximum number of re-training rounds $E_r=100$ in \textsc{Retrain} to not slow down model upload speed. Therefore, \textsc{Retrain} terminates when the number of re-training rounds reaches $E_r$, even if $Acc(w^+_{G(t+1)}, \mathrm{WM}_{w_G}) < th$. 

\ourmethod starts with \aggt 
executing \textsc{Pretrain}. Then \agg\ executes \textsc{Retrain} and sends $w^+_{G(t+1)}$ to $C_{sub}$ in each $t$. 
By starting federated learning with a global model already converged to the watermark, we can decrease the number of re-training rounds required to reach high $Acc(w^{+}_{G(t)}, \mathrm{WM}_{w_{G}})$ satisfying~\ref{req:p_computation}. \ourmethod ensures high enough $Acc(w^{+}_{G(t)}, \mathrm{WM}_{w_{G}})$ to enable demonstration of ownership at anytime, satisfying~\ref{req:w_ownership}. \ourmethod also ensures that $w^{+}_{G(t)}$ and $w^{+}_{adv(t)}$ converge for both the watermark and the actual task, satisfying~\ref{req:p_utility}. In addition, \ourmethod\ does not require any client's training data to operate, since \ourmethod\ re-trains $w^{+}_{G(t)}$ using only the watermark but no other data samples. Therefore, it satisfies the requirement~\ref{req:w_data_independence}. 

\ourmethod is executed after the aggregation step of federated learning, and \textbf{is independent of the aggregation method}. Therefore, any other robust aggregation method such as Krum~\cite{blanchard2017machine}, trimmed mean or median~\cite{yin2018byzantine} can be easily combined with \ourmethod.
While \ourmethod\ can use any existing watermark set, we introduce \ourpattern, a novel way to generate (training) data-independent watermarks.
\subsection{\ourpattern}\label{ssec:watermarkgeneration}
\begin{figure*}[t]
    \centering
       \subfloat[Clean\label{fig:originalimage}]{\includegraphics[width=0.069\linewidth]{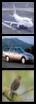}}
       \quad
        \subfloat[\ourpattern\label{fig:WMuPattern}]{\includegraphics[width=0.20\linewidth]{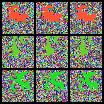}}
        \quad
        \subfloat[Embedded C.\label{fig:WMrPattern}]{\includegraphics[width=0.20\linewidth]{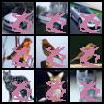}}
        \quad
        \subfloat[unRelate\label{fig:WMunRelate}]{\includegraphics[width=0.20\linewidth]{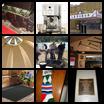}}
        \quad
        \subfloat[unStruct\label{fig:WMunStruct}]{\includegraphics[width=0.20\linewidth]{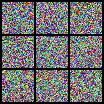}}
       \caption{Original images (a) in CIFAR10 dataset for classes airplane, automobile and bird as well as sample of watermark sets \ourpattern\ (b), Embedded Content~\cite{zhang2018protecting} (c), unRelate~\cite{adi2018turning,zhang2018protecting} (d) and unStruct~\cite{rouhani2018deepsigns} (e) constructed for these three classes.} \label{fig:watermarkexamples}   
\end{figure*}

Inspired by the prior work~\cite{adi2018turning,zhang2018protecting} and assuming that $O$ has no access to the training data in large-scale federated learning applications, we propose \ourpattern: adding specific patterns to images containing only noise.
While creating \ourpattern, we first generate different images using Gaussian noise. Then, each image is embedded with a certain pattern and labeled with a class that is related to the original task. Each class has a different pattern and every pattern is unique in terms of color, shape, orientation and position. We assume that $O$ knows the dimensions of the model's inputs and outputs, since $O$ selects the model architecture according to these specifications. Therefore, this information is enough to construct \ourpattern. Figure~\ref{fig:WMuPattern} illustrates samples of \ourpattern\ generated for colored images. 

\ourpattern is \textbf{independent of training data} and suitable for federated learning. This property satisfies the requirement \ref{req:w_data_independence}. Additionally, \ourpattern is \textbf{easy to learn}. Using the same pattern for each class helps $w_G$ to converge and overfit to \ourpattern, since samples of each class include features that are easy to learn and memorize. Therefore, this property helps satisfying \ref{req:p_communication} and \ref{req:p_computation}.
\section{Experimental Setup}\label{sec:expsetup}

\subsection{Datasets and Models}\label{sssec:datasets} Considering that reaching a high accuracy in federated learning is 
very challenging, we replicated image classification tasks in~\cite{mcmahan2016communication} to provide baseline models with a sufficiently high test accuracy. We choose MNIST~\cite{deng2012mnist} and CIFAR10~\cite{krizhevsky2014cifar} as datasets since prior work~\cite{bagdasaryan2018backdoor,li2019fedmd,mcmahan2016communication} has obtained good models with these datasets using FedAvg. 
\removetext{MNIST consists of 60,000 grayscale images of 10 digits as training data with an additional test set of 10,000 grayscale images. CIFAR10 includes 50,000 color images labeled with 10 categories as training data, and 10,000 color images as test data. }
For MNIST, we use a 5-layer convolutional neural network~\cite{juuti2019prada}. For CIFAR10, we use VGG16~\cite{simonyan2014very}\footnote{VGG16 has no batch normalization layer that uses training data statistics. Therefore, it is a suitable model for federated learning.} which is an off-the-shelf complex DNN model trained over the ImageNet dataset~\cite{deng2009imagenet}. 
For the federated learning setup, we use 100 \client s participating the training and select 10 \client s randomly as $C_{sub}$ in each $t$. 
We repeat our experiments for different number of local passes $E_c=\{1, 5, 10, 20\}$ used by each client. 
We distribute the training data over clients in IID fashion. Each \client\ receives 600 and 500 training images for MNIST and CIFAR10, respectively. We measure the test accuracy using the test subset of datasets, assuming a similar setup as 
in~\cite{hard2018federated}: the model is evaluated at different client devices not involved in the federated learning process. 
Details about the training configuration settings can be found in Appendix~\ref{ssec:design}.

We compare \ourmethod\ to other methods presented in~\ref{ssec:wmoverview} and to baseline models that do not use a watermarking scheme. Baseline models are constructed using FedAvg and we use the tuple of $\{E_c, E_a\}$ to define different baseline models.
\removetext{We also obtain Pre-embedded models, where $O$ embeds his watermark $\mathrm{WM}_{w_G}$ by training $w_{G(0)}$ with it. $O$ sends parameters of the watermarked model $w^{+}_{G(0)}$ to the cloud server before the federated learning starts.}
We achieve $99\%$ test accuracy on MNIST in all 4 baseline models. For CIFAR10, we achieve $85\%$ test accuracy similar to~\cite{mcmahan2016communication}. 

\subsection{Watermark Sets}\label{ssec:watermarkimplementation}

For generating \ourpattern\ 
we choose 100 for the size of the watermark set since it is sufficient to provide a high confidence $>1 - 2^{-64}$ during demonstration of ownership~\cite{adi2018turning,szyller2019dawn} while not degrading the overall performance of the model. In both MNIST and CIFAR10, each class contains 10 watermark samples. In order to construct \ourpattern, we first generate 100 images with Gaussian noise, then randomly create 10 different patterns and label each pattern with a different class. Finally, each noisy image is combined with a random pattern. Figure~\ref{fig:WMuPattern} shows samples of \ourpattern\ generated for the CIFAR10 task. 

\newtext{We also compare the performance of \ourpattern\ to other state-of-the-art watermark generation methods: Embedded Content~\cite{zhang2018protecting}, unRelate~\cite{adi2018turning,zhang2018protecting} and unStruct~\cite{rouhani2018deepsigns}. Figure~\ref{fig:watermarkexamples} shows samples of these watermarks generated for the CIFAR10 task. Details about these watermark sets are given in Appendix~\ref{app:generatewmsets}.}
\section{Evaluation}\label{sec:wmevaluation}
We evaluate the performance of \ourmethod\ and \ourpattern\ using the requirements defined in Section~\ref{ssec:requirements}. We compare the performance of models watermarked using \ourmethod\ to Pre-embedded models and baseline models without watermark (see Section~\ref{sec:expsetup}). We also analyze how well \ourpattern\ performs by comparing it to Embedded Content~\cite{zhang2018protecting}, unRelate~\cite{adi2018turning,zhang2018protecting} and unStruct~\cite{rouhani2018deepsigns} (see Figure~\ref{fig:watermarkexamples}).
Further evaluation with additional experimental results can be found in Appendix~\ref{app:additionalexperimentalresults}.

\subsection{\ref{req:w_ownership} Demonstration of Ownership}\label{ssec:wmdemonstration}

\ref{req:w_ownership} requires $\textsc{Verify}$ to return $True$ at any $t$. We explained in Section~\ref{sec:method} that $\mathrm{WM}_{w_{G}}$ is secret, only known by the model owner and \agg. It can be registered in a public bulletin as proposed in~\cite{adi2018turning}. Thus, \ref{req:w_ownership} is satisfied if $Acc(w^+_{G(t)}, \mathrm{WM}_{w_{G}}) \geq T_{acc}$. Using the formula defined in~\cite{szyller2019dawn}, we compute that a reliable demonstration of ownership with confidence $>1 - 2^{-64}$, given a watermark set of size 100 for a 10-classes classifier (MNIST and CIFAR10 models), is provided by $T_{acc} = 47\%$.

Table~\ref{tb:accuracy} 
shows that Pre-embedding achieves on average $Acc(w^+_{G(E_a)}, \mathrm{WM}_{w_{G}})=27\%$ for MNIST and $15\%$ for CIFAR10. Watermarks embedded via Pre-embedding are not resilient to re-training. In contrast, \ourmethod achieves high watermark accuracy ($\sim 99.0\%$) at the last aggregation round in both tasks and
$\textsc{Verify}$ returns $True$ before $w_{G}$ starts to improve (after 10 aggregation rounds on average). We can successfully embed all four types of watermark sets long before the global model converges. Therefore, in the remaining experiments, we only evaluate \ourmethod with different watermark sets. 

\begin{table}[t]
\caption{Average watermark accuracy (over 4 different watermarks) at the final aggregation round $E_a$. 
\ourmethod satisfies \ref{req:w_ownership} while Pre-embedding does not: $Acc(w^+_{G(E_a)}, \mathrm{WM}_{w_{G}}) < 47\%$.}\label{tb:accuracy}
\centering
\resizebox{\columnwidth}{!}{
\begin{tabular}
                       {c|c|c|c|c}
\hline 
& \multicolumn{2}{|c|}{$Acc(w^+_{G(E_a)}, \mathrm{WM}_{w_{G}})$ for \textbf{MNIST}} & \multicolumn{2}{|c}{$Acc(w^+_{G(E_a)}, \mathrm{WM}_{w_{G}})$ for \textbf{CIFAR10}} \\ 
$\{E_c, E_a\}$ & \hfil Pre-embedding & \hfil {\ourmethod}& \hfil Pre-embedding & \hfil {\ourmethod}
\\ \hline
\small $\{1, 250\}$ & \hfil 24.00& \hfil  99.00&\hfil 15.00 & \hfil 99.00 \\ 
\small $\{5, 200\}$ & \hfil 30.00&  \hfil 99.00& \hfil 14.00 &\hfil 99.50\\ 
\small $\{10, 150\}$ & \hfil 22.75& \hfil 98.50& \hfil 15.00 & \hfil 99.00\\
\small $\{20, 100\}$ & \hfil 31.00& \hfil 98.75& \hfil 16.00 & \hfil 99.75\\ \hline
\end{tabular}}
\end{table}

\subsection{\ref{req:w_robustness} Robustness}\label{ssec:wmrobustness}
\ref{req:w_robustness} states that embedded watermarks should be robust against post-processing watermark removal techniques. For evaluating the robustness, we use three state-of-the-art defenses against backdooring. 
Fine-tuning~\cite{yosinski2014transferable} and pruning~\cite{han2015learning,he2017channelpruning} 
are generic watermark removal techniques that do not require knowledge of the trigger pattern or the watermarking method. Both techniques utilize some clean dataset (e.g., subset of a training set) and re-train the model in order to remove the watermark. 
Neural Cleanse~\cite{wang2019neural} is a technique to detect, reverse-engineer and subsequently remove potential backdoors from DNN models, and it can be used for watermark removal. To evaluate these techniques, we set the acceptable utility drop of 5 percentage point (pp): the watermark must be removed while keeping the test accuracy degradation less than 5pp. 


\subsubsection{Fine-tuning attack}\label{sssection:finetuning}

\begin{figure}[t]
    \centering
        \subfloat
        {\includegraphics[width=0.90\columnwidth]{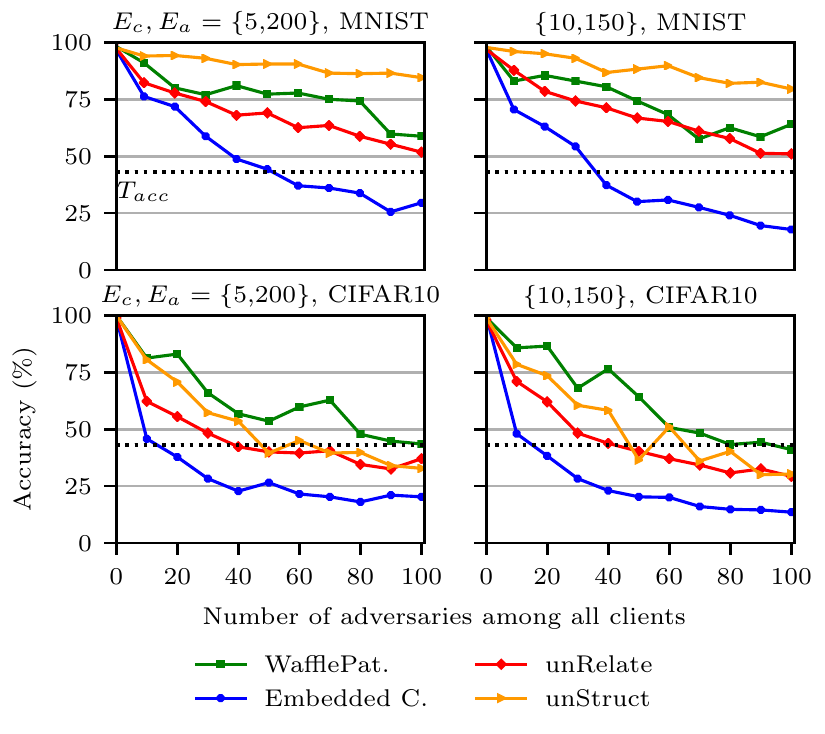}}
       \caption{Comparison of watermark accuracy for MNIST and CIFAR10 at different $\{E_c, E_a\}$ tuples when fine-tuning is implemented by an increasing number of malicious clients.}\label{fig:finetuningresults}
\end{figure}

\begin{table}[t]
\caption{Test and watermark (WM) accuracy (\%) for different numbers of fine-tuning epochs on watermarked MNIST and CIFAR10 models using \ourpattern at different $\{E_c, E_a\}$ tuples. 1 adversary out of 100 clients.}\label{tb:finetuningtrendMNISTCIFAR}
\centering
\resizebox{\columnwidth}{!}{
\begin{tabular}
                       {c|c|c|c|c|c|c|c|c}
\hline
& \multicolumn{4}{|c|}{MNIST} & \multicolumn{4}{|c}{CIFAR10} \\  \hline
\multicolumn{1}{c|}{\footnotesize $\{E_c, E_a\}$} &  \multicolumn{2}{c|}{ $\{5,200\}$} &  \multicolumn{2}{c|}{$\{10,150\}$} &  \multicolumn{2}{c|}{$\{5,200\}$} &  \multicolumn{2}{c}{$\{10,150\}$} \\
 \hline
epoch &\hfil Test & \hfil WM& \hfil Test & \hfil WM & \hfil Test & \hfil WM & \hfil Test & \hfil WM \\ \hline

 0    & \hfil 99.0& \hfil 99.0& \hfil 98.9 & \hfil  100.0 & \hfil 85.6 &\hfil  100.0 & \hfil  85.8 & \hfil 99.0 \\
 
20   & \hfil 98.9& \hfil 98.8& \hfil 98.7  & \hfil 98.8 & \hfil 85.5 & \hfil 99.8 & \hfil   85.5 & \hfil 96.8 \\

40   & \hfil 98.9 & \hfil 98.5 & \hfil 98.7  & \hfil 98.8 & \hfil 85.5 & \hfil 99.8 & \hfil 85.5 & \hfil 96.2 \\

60   & \hfil 98.9 & \hfil 98.2 & \hfil 98.7  & \hfil 98.2 & \hfil 85.5 & \hfil 99.8 & \hfil 85.6 & \hfil 96.0 \\

80   & \hfil 98.9 & \hfil 98.2 & \hfil 98.7  & \hfil 98.0 & \hfil  85.5 & \hfil 99.8 & \hfil 85.6  & \hfil 96.0 \\

100 & \hfil 98.9 & \hfil 98.0 & \hfil 98.7  & \hfil 98.0 & \hfil 85.5 & \hfil 99.8 & \hfil 85.6  & \hfil 96.2 \\ \hline
 \end{tabular}}
\end{table}

We tested fine-tuning attack using an increasing number of malicious clients combining their local datasets. \newtext{Figure~\ref{fig:finetuningresults} and~\ref{fig:finetuningresults2} (Appendix~\ref{app:additionalexperimentalresults}) show that \ourpattern is the most resilient watermark to fine-tuning for CIFAR10 in all experiments while it is the second most resilient watermark for MNIST when $ \{E_c, E_a \}=\{5, 200\}$ and $\{10, 150\}$.} Table~\ref{tb:finetuningtrendMNISTCIFAR} and~\ref{tb:finetuningtrendMNISTCIFAR2} (Appendix~\ref{app:additionalexperimentalresults})
provide a detailed evaluation of resilience according to the number of fine-tuning epochs run by $adv$. Tables show that \ourpattern is resilient to fine-tuning even for a large number of fine-tuning epochs.

\subsubsection{Pruning attack}\label{sssection:pruning}
\begin{figure}[t]
    \centering
         \subfloat
         {\includegraphics[width=0.90\columnwidth]{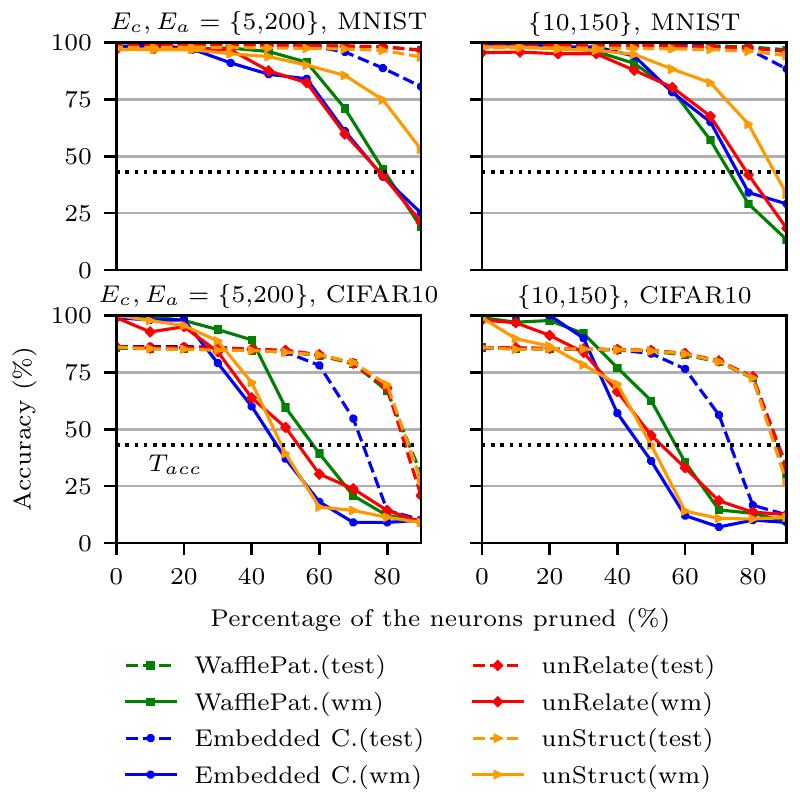}}
       \caption{Comparison of the test and watermark (wm) accuracy for MNIST and CIFAR10 at different $\{E_c, E_a\}$ tuples when the pruning attack is implemented by one adversary.}\label{fig:pruningresultsPR}  
\end{figure}

We implemented parameter pruning proposed in~\cite{han2015learning} followed by fine-tuning by one $adv$ using different pruning rates. We chose the pruning technique in~\cite{han2015learning} that first removes a number of connections with magnitudes close to zero from the dense model and obtains a more sparse model. Then, it re-trains this sparse model to recover the test accuracy. \newtext{Figure~\ref{fig:pruningresultsPR} and~\ref{fig:pruningresultsPR2} 
Appendix~\ref{app:additionalexperimentalresults}) show that \ourpattern is the most resilient watermark for CIFAR10, since the watermark accuracy decreases below $T_{acc}$ when a higher percentage of neurons removed from CIFAR10 models compared to other watermark types. Although it is not the most resilient in MNIST,  Figure~\ref{fig:pruningresultsPR} shows that if the adversary removes more than 70\% of the neurons from MNIST models (respectively 50\% in CIFAR10 models), test accuracy starts to decline.}
Therefore, even though $adv$ can evade demonstration of ownership, i.e., $\textsc{Verify}(w_{adv}, \mathrm{WM}_{G})\rightarrow False$, the pruned model shows a drop in test accuracy larger than 5pp compared to the original $w_{adv}$. We also measured that \ourpattern satisfies~\ref{req:w_robustness} if the proportion of malicious clients is smaller than 
40\%.

\subsubsection{Neural Cleanse attack}
Neural Cleanse first finds potential small trigger patterns needed to misclassify all inputs as a specific label and repeats this step for all classes. Then, it outputs an anomaly index based on this analysis. If the anomaly index is above 2.0, the model is considered backdoored. In the case of a suspected backdoor, Neural Cleanse also marks any class that is likely to be infected. We found that adversaries cannot detect the presence of watermarks in MNIST (Table~\ref{fig:reversedtriggers}, Appendix~\ref{app:additionalexperimentalresults}). On the other hand, $adv$ can recognize that the model is watermarked in CIFAR10. However, it can only mark up to 3 possibly infected classes, while all ten classes are infected by our watermarks. Moreover, the reversed triggers of marked classes look completely different from the original patterns, as shown in Figure~\ref{fig:reversedtriggers} (Appendix~\ref{app:additionalexperimentalresults}). 

We also implemented patching-via-unlearning~\cite{wang2019neural}, a mitigation technique using reversed triggers obtained via Neural Cleanse. This method requires relaxing some assumptions of our adversary model and leaking information about the triggers to $adv$. We assumed that $adv$ is aware of the presence of watermarks in $w_{adv}$ and it knows that the watermark set includes different, distinctive patterns for each class. 
Performing this attack, a single $adv$ cannot remove the watermark from MNIST model and at least 40\% of clients should be malicious for a successful removal. (Figure~\ref{fig:ncleanseresultsPR}, Appendix~\ref{app:additionalexperimentalresults}).
For CIFAR10, one $adv$ might evade the verification but the performance drop for watermarked models using \ourpattern is over 5pp after patching as shown in Table~\ref{tb:neuralcleansetrend}. If less than 10\% of clients are malicious, they cannot recover the performance of $w_{G}$ and \ourpattern satisfies~\ref{req:w_robustness}.

\begin{table}[t]
\caption{Robustness against NeuralCleanse patching via unlearning. Test and watermark (WM) accuracy (\%) for CIFAR10 models watermarked using \ourpattern. }\label{tb:neuralcleansetrend}
\resizebox{\columnwidth}{!}{
\begin{tabular}{
                       c|cc|cc|cc|cc}
 \hline
{\footnotesize $\{E_c, E_a\}$} &  \multicolumn{2}{c|}{ $\{1,250\}$} &  \multicolumn{2}{c|}{$\{5,200\}$} &  \multicolumn{2}{c|}{$\{10,150\}$} &  \multicolumn{2}{c}{$\{20,100\}$} \\
 \hline
\# of adv. &  \hfil Test  &  \hfil WM  &  \hfil Test  &  \hfil WM  &  \hfil Test  &  \hfil WM  &  \hfil Test  &  \hfil WM \\ \hline
 0 (Baseline) & 85.7& 100.0 & 85.6  & 100.0 & 85.8 & 99.0  & 85.6  & 100.0\\
 1   & 74.0 & 27.2 & 75.9   & 25.2 & 75.6 & 28.0  & 76.2  & 35.0 \\
 2   & 78.8 & 29.5 & 73.4   & 31.8 & 72.6 & 26.0  &  74.4 & 26.5 \\
 5   & 78.1 & 29.8 & 80.5   & 26.0 & 78.5 & 31.2  & 78.8  & 24.8\\
 10 & 80.2 & 27.5 & \textcolor{red}{80.8}   & \textcolor{red}{34.5} & \textcolor{red}{80.9} & \textcolor{red}{20.0}  &  78.4 & 38.8\\
 20 & 79.3 & 36.5 & \textcolor{red}{81.6}   & \textcolor{red}{30.2} & \textcolor{red}{81.5} & \textcolor{red}{25.2}  &  \textcolor{red}{80.8} & \textcolor{red}{36.5}\\ 
 30 & 80.6 & 17.8 & \textcolor{red}{82.4}   & \textcolor{red}{26.5} & \textcolor{red}{81.0} & \textcolor{red}{34.8}  &  \textcolor{red}{82.0} & \textcolor{red}{37.5} \\ 
 40 & \textcolor{red}{83.3} & \textcolor{red}{30.8} & \textcolor{red}{81.8}   & \textcolor{red}{27.8} & \textcolor{red}{81.1} & \textcolor{red}{40.2}  & \textcolor{red}{81.5}  & \textcolor{red}{30.0}\\ \hline
 \end{tabular}}
\end{table}

\subsection{\ref{req:p_utility} Model Utility}\label{ssec:modelutility}
According to \ref{req:p_utility}, watermarking should not degrade the test accuracy of the converged model, i.e., $w^{+}_{G(t=E_a)}$ and $Acc(w^{+}_{G(t=E_a)}, D_{test}) \approx Acc(w_{adv(t=E_a)}, D_{test})$. For evaluating utility, we measured the test accuracy of watermarked federated learning models using four different watermark sets, and compared it to non-watermarked baseline models. Table~\ref{tb:testaccuracy} presents the test accuracy of baseline and watermarked models using different $\{E_c, E_a\}$ combinations. Results show that \ourpattern reaches a sufficient test accuracy similar to baseline models ($\leq 0.2pp$ for MNIST and $\leq 0.7pp$ for CIFAR10.) and does not degrade the test accuracy as much as unStruct in MNIST or Embedded Content in CIFAR10. \ourpattern satisfies \ref{req:p_utility}. 

\subsection{\ref{req:p_communication}-\ref{req:p_computation} Communication and computational overhead}\label{ssec:evaloverhead}
\ref{req:p_communication} and \ref{req:p_computation} state that both the watermarking procedure and the watermark set should not increase the communication overhead and incur minimal additional computation. 
\ourmethod increases computation by re-training the global model in each $t$. Different watermark sets may be easier/harder to learn and incur additional communication (aggregation rounds) for the model to converge for both the watermark set and its main task: satisfying both \ref{req:w_ownership} and \ref{req:p_utility}.

\begin{figure}[t]
     \centering
        \subfloat
    {\includegraphics[width=0.85\columnwidth]{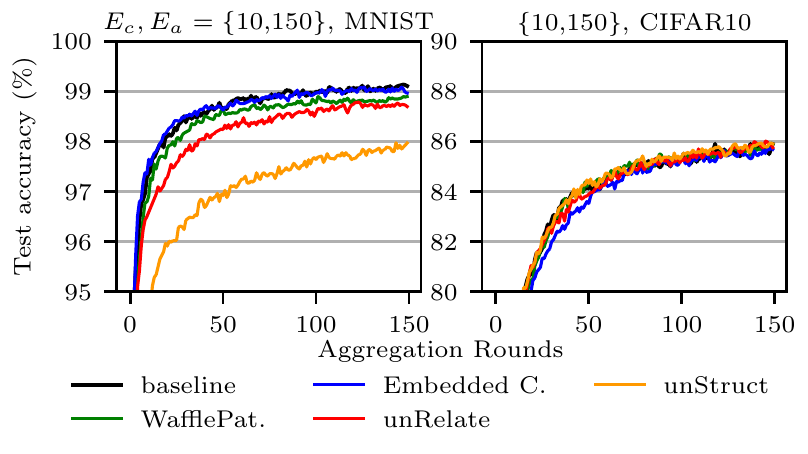}}
    \caption{Progression of the test accuracy for MNIST and CIFAR10. Baseline and watermarked models using different watermark sets are shown in the figure.}\label{fig:commoverhead} 
\end{figure}

We evaluated the communication overhead by measuring the increase in test accuracy of watermarked models according to $E_{a}$. Reaching a high test accuracy in a smaller $E_{a}$ means that the model can be trained with minimal communications. Figure~\ref{fig:commoverhead} and~\ref{fig:commoverhead2} (Appendix~\ref{app:additionalexperimentalresults}) compares the test accuracy progression of non-watermarked baseline models and watermarked models
using different watermark sets. In MNIST, the test accuracy quickly converges in baseline models and all watermarked models except for unStruct. In CIFAR10, Embedded Content converges slower than other watermark sets and requires more aggregation rounds for obtaining a performance similar to baseline models. In both cases, \ourpattern satisfies \ref{req:p_communication}.

We also calculated the computational overhead by dividing the total number of retraining rounds to the total number of local passes performed by clients. Table~\ref{tb:computationaloverhead} gives the computational overhead in \ourmethod using different watermarks. 
While unStruct has the lowest computational overhead, it requires twice as much aggregation rounds compared to other models to reach $99\%$ test accuracy in  MNIST. 
\ourpattern\ usually needs fewer retraining rounds than other watermark sets, since it contains similar features that can be learned easily for each class. Therefore, \ourpattern satisfies \ref{req:p_computation}.

\begin{table}[t]
\caption{The test accuracy (\%) (at $t=E_a$) of watermarked models using different watermark sets.}\label{tb:testaccuracy}
\centering
\resizebox{\columnwidth}{!}{
\begin{tabular}
                       {c|c|c|c|c|c}
\hline 
 $\{E_c, E_a\}$& \multicolumn{5}{|c}{Watermark pattern}   \\ \hline
MNIST & Baseline &  \ourpattern & Embedded C. & unRelate & unStruct \\ \hline
 \small $\{1,250\}$  & \hfil 98.97 & \hfil 98.88  & \hfil \textcolor{greena}{99.05} & \hfil 98.92 & \hfil \textcolor{red}{97.59}   \\ 
 \small $\{5,200\}$   & \hfil 98.91  & \hfil 98.94 & \hfil \textcolor{greena}{98.98} & \hfil 98.79 & \hfil  \textcolor{red}{98.13} \\ 
 \small $\{10,150\}$ & \hfil 99.11 & \hfil \textcolor{greena}{99.06}  & \hfil 98.97  & \hfil  \textcolor{greena}{99.06} & \hfil  \textcolor{red}{97.97} \\ 
 \small $\{20,100\}$& \hfil 99.02 & \hfil 98.85 & \hfil \textcolor{greena}{98.97} & \hfil 98.79 & \hfil  \textcolor{red}{97.77}    \\ \hline
CIFAR10 & Baseline &  \ourpattern & Embedded C. & unRelate & unStruct \\ \hline
 \small $\{1,250\}$  & \hfil 86.27 &  \hfil 85.70 & \hfil \textcolor{red}{85.19} & \hfil 85.81 & \hfil \textcolor{greena}{86.53}   \\ 
 \small $\{5,200\}$  & \hfil 86.24 & \hfil \textcolor{red}{85.61} & \hfil 86.21 & \hfil \textcolor{greena}{86.25} & \hfil 85.99  \\
 \small $\{10,150\}$& \hfil 85.90 & \hfil 85.89 & \hfil \textcolor{red}{85.69} & \hfil 85.76  & \hfil \textcolor{greena}{85.91}  \\ 
 \small $\{20,100\}$& \hfil 85.85 & \hfil 85.67 & \hfil \textcolor{red}{85.47}  & \hfil \textcolor{greena}{85.74} & \hfil 85.72    \\  \hline
\end{tabular}}
\end{table}

\begin{table}[t]
\caption{Average computational overhead (\%) incurred by the retraining rounds in watermarked models using different watermark sets.}\label{tb:computationaloverhead}
\centering
\resizebox{\columnwidth}{!}{
\begin{tabular}{ p{\dimexpr 0.18\linewidth-2\tabcolsep}
                       | p{\dimexpr 0.20\linewidth-2\tabcolsep}
                       | p{\dimexpr 0.20\linewidth-2\tabcolsep}
                       | p{\dimexpr 0.20\linewidth-2\tabcolsep}
                       | p{\dimexpr 0.20\linewidth-2\tabcolsep}}
\hline
\multicolumn{1}{c|}{Dataset} &  \multicolumn{1}{|c|}{\textbf{\ourpattern}} & \multicolumn{1}{|c|}{Embedded C.} & \multicolumn{1}{|c|}{unRelate} & \multicolumn{1}{|c}{unStruct} \\ \hline 
MNIST & \hfil 3.06  & \hfil 2.02   & \hfil \textcolor{red}{10.39}  & \hfil \textcolor{greena}{0.91} \\ 
CIFAR10 & \hfil 2.97 & \hfil 5.72  & \hfil \textcolor{red}{6.10} & \hfil   \textcolor{greena}{1.47} \\ \hline
\end{tabular}}
\end{table}

\subsection{Evasion of Verification}

In addition to methods to recover and remove watermarks, $adv$ can also attack the verification mechanisms~\cite{adi2018turning,szyller2019dawn} used to demonstrate model ownership. Even though verification is not in the scope of our paper, we discuss possible attacks.

$adv$ may try to \textit{evade verification} by detecting queries for watermark samples as out-of-distribution (OOD) samples. We implemented and tested this attack using the threshold-based detector model introduced in~\cite{li2019prove} for CIFAR10. This method is shown~\cite{li2019prove} to be strong enough to evade verification against backdoor-based watermarking methods~\cite{adi2018turning,guo2018watermarking} with a negligible false positive rate (FPR). As a possible watermark set, we use a subset of the TinyImageNet\footnote{\url{https://tiny-imagenet.herokuapp.com}}, which is similar to unRelate. $adv$ trains the detector with both its training data which represents in-distribution data, and TinyImageNet subset representing OOD data. We investigate two different scenarios: 1) Each client including adversaries, has a balanced, IID dataset as defined in the adversary model, and 2) a more realistic scenario where clients as well as adversaries have non-IID, unbalanced datasets~\cite{mcmahan2016communication}. In both scenarios, CIFAR10 models are watermarked 
with \ourpattern where $\{E_{c_{1}}, E_{a_{1}} \}=\{1, 250\}$, $\{E_{c_{2}}, E_{a_{2}} \}=\{1, 450\}$ and $w_{adv}^2$ has 82.5\% test accuracy so the two models have similar performance. Table~\ref{tb:evasionFPR} reports both the lowest FPR calculated over multiple thresholds and the true positive rate (TPR, the ratio of watermark samples correctly identified as OOD to the watermark set) at that FPR. As can be seen from the table, \ourmethod watermark verification could be evaded if $adv$ has high quality IID data. However, \ourmethod is resilient to evasion in a non-IID setting. In the non-IID scenario, verification can be evaded with 5\% FPR only if more than 50\% clients are adversaries and share their datasets for training the detector. Moreover, since $adv$ having only a limited training dataset might choose poor OOD data, this affects the detection performance and increases FPR to a degree such that the resulting prediction model is unusable~\cite{song2020critical}. Therefore, evasion of verification is not a concern for real-world client-server federated learning with non-IID data.
\begin{table}[t]
\caption{Evasion of verification in both IID and non-IID settings. The lowest false positive rate (FPR) and true positive rate (TPR) calculated at that FPR is reported for watermarked CIFAR10 models using \ourpattern.}\label{tb:evasionFPR}
\centering
\begin{tabular} {c|cc|cc}
\hline
& \multicolumn{2}{c}{IID setting} & \multicolumn{2}{|c}{non-IID setting} \\ \hline
\# of adv. & TPR & (lowest) FPR & TPR & (lowest) FPR \\ \hline
1 &   64.0 &  0.8 & 89.5 & 53.0\\ 
2 &   78.7 &  1.3 & 80.8 & 39.3\\
5 &   88.0 &  1.6 & 92.2 & 22.9\\
10 & 94.7 &  2.5 & 90.8 & 19.7\\
20 & 90.0 &  1.1 & 91.8 & 7.0\\
30 & 96.5 &  1.6 & 88.0 & 15.3\\
40 & 81.0 &  1.0 & 91.8 & 6.8\\ 
50 & 80.0 &  0.6 & 84.0 & 4.8\\ \hline
\end{tabular}
\end{table}
\section{Discussion and Takeaways}\label{ssec:discussion}
 
\begin{figure}[t]
    \centering{\includegraphics[width=0.62\columnwidth]{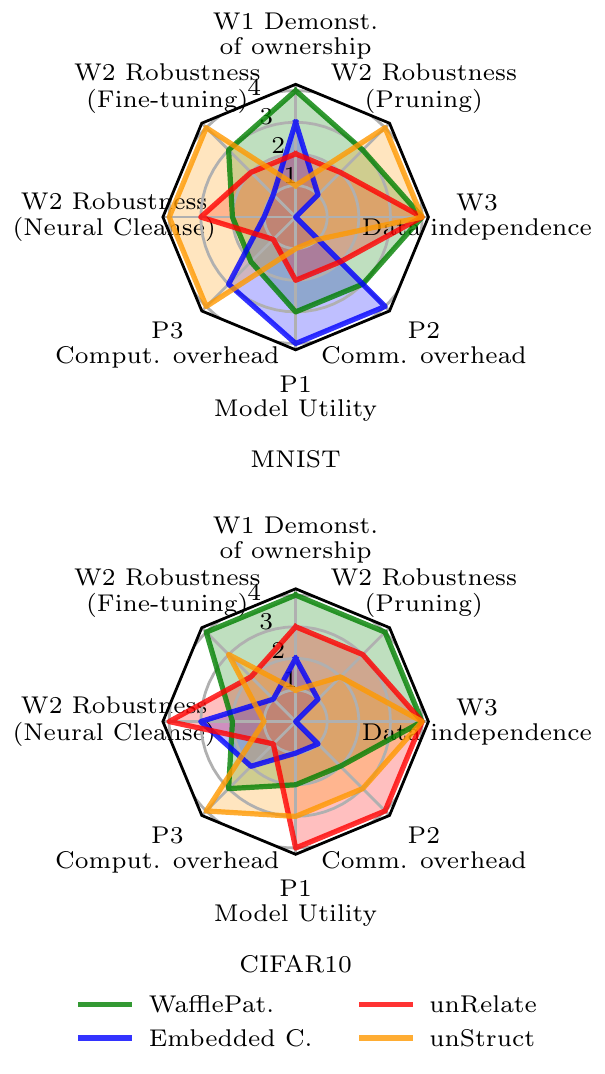}}
       \caption{\newtext{Trade-off between meeting the requirements in Section~\ref{ssec:requirements} for different watermark sets. All watermarks are ranked between 1-4 (the higher, the better) based on the results obtained from all experiments.}} \label{fig:radarchart}   
\end{figure}

In contrast to existing Pre- and Post-embedding techniques, \ourmethod\ meets~\ref{req:w_ownership} enabling to reliably demonstrate ownership of $w_{G}$ at any aggregation round. 
\newtext{Figure~\ref{fig:radarchart} summarizes the overall effectiveness of different watermark types, and it is evident that \ourpattern gives the best trade-off (big area and never the worst one) considering all six requirements.}


\removetext{
\begin{table*}[t]
\caption{Ranking (1-4) for meeting the requirements in Section~\ref{ssec:requirements} for different watermark sets for MNIST (left) and CIFAR10 (right).}\label{tb:wmsetssummary}
\centering
\begin{tabular} {
                       |c|c|c|c|}
\hline 
\textbf{MNIST/CIFAR10}&  \hfil \ourpattern & \hfil Embedded C. & \hfil unRelate & \hfil unStruct \\ \hline
~\ref{req:w_ownership} Demonstration of ownership & \hfil 1/1 & \hfil 2/3 & \hfil 3/2 & \hfil 3/3\\ \hline
~\ref{req:w_robustness} Robustness (Fine-tuning) & \hfil 2/1 & \hfil 4/4 & \hfil 3/3 & \hfil 1/2\\ \hline
~\ref{req:w_robustness} Robustness (Pruning) & \hfil 2/1 & \hfil 4/4 & \hfil 3/2 & \hfil 1/3\\ \hline
~\ref{req:w_robustness} Robustness (Neural Cleanse) & \hfil 3/3 & \hfil 4/2 & \hfil 2/1 & \hfil 1/4\\ \hline
~\ref{req:w_data_independence} Data Independence & \hfil Yes/Yes & \hfil No/No & \hfil Yes/Yes & \hfil Yes/Yes\\ \hline
~\ref{req:p_utility} Model utility & \hfil 2/3 & \hfil 1/4 & \hfil 3/1 & \hfil 4/2\\ \hline
~\ref{req:p_communication} Comm. overhead & \hfil 2/3 & \hfil 1/4 &\hfil  3/1 & \hfil 4/2\\ \hline
~\ref{req:p_computation} Comput. overhead & \hfil 3/2 & \hfil 2/3 &\hfil  4/4 & \hfil 1/1\\ \hline
\end{tabular}
\end{table*}}

\ourmethod is also resilient to almost all watermark removal attacks if less than 40\% clients are malicious, as summarized in Table~\ref{tb:overallresilience}. While Neural Cleanse requires only 10\% malicious clients to be successful against the CIFAR10 model, adversaries would have to share their datasets and collaborate with each other to recover the triggers.
The collaboration between adversaries in a distributed attack is different from a sybil attack in our case. Creating multiple identities does not improve removal and evasion attacks since the performance of these attacks is directly related to the quantity and the diversity of data held by the adversary. 
Collaboration is impractical since adversaries need to reveal their highly-sensitive dataset to other untrustworthy parties.
We also empirically evaluated that watermark removal attacks fail to decrease watermark accuracy or recover test accuracy, if performed in a federated learning setup without sharing their datasets. On the other hand, if an adversary holds more than 10\% of the whole training data, it could successfully remove watermarks from $w_{adv}$ without decreasing the model performance. This scenario is not typical in large-scale distributed learning settings, where there are a very large number of clients, each holding a tiny fraction of the overall training data. 
We conclude that under reasonable assumptions, \ourmethod is resilient to distributed attacks of several malicious federated learning clients.

\begin{table}[t]
\caption{Minimum ratio of malicious clients required to evade \ourmethod.}\label{tb:overallresilience}
\centering
\resizebox{0.95\linewidth}{!}{
\begin{tabular}{c|c|c}
 \hline 
 \textbf{Attack type}  & \textbf{MNIST} & \textbf{CIFAR10} \\ 
 \hline 
 Fine-tuning~\cite{yosinski2014transferable} & 50\% & 50\% \\ 
 \hline 
 Pruning~\cite{han2015learning} & 40\% & 40\%  \\ 
 \hline 
 Neural Cleanse, patching-via-unlearning~\cite{wang2019neural} & 40\% & 10\% \\ 
 \hline
Evasion of verification with OOD detection~\cite{li2019prove} & - & 50\% \\ 
 \hline 
 \end{tabular}}  
 \end{table}

Adversaries can also try to cast model ownership into doubt by embedding its own watermark into the model: \textit{ownership piracy}. This issue can be addressed by registering any watermark into a time-stamped bulletin (e.g., blockchain) for it to be valid~\cite{adi2018turning,szyller2019dawn}. We can enforce our watermark to be registered together with some private artifacts of the model before federated learning starts  as a proof of authenticity. 
Such a private artifact can be the architecture of the model (number of layers, number of neurons, type of activation functions, etc.) since all these parameters can provide a large entropy and they are unknown to any federated learning client before they receive the initial model.

Finally, a model stealing attack can be performed against $w_{adv}$ by $adv$ itself to create a surrogate model without watermark or embed its own watermark~\cite{shafieinejad2019robustness,szyller2019dawn}. Model stealing typically requires a large amount of data (more than specified by our adversary model) and it causes accuracy drops larger than 5pp~\cite{atli2019extraction}. Thus, we consider these attacks impractical in our adversary model.

\newtext{We assume that each client's training data $D_{c_i}$ is 
IID (Section~\ref{ssec:adversarymodel}), and based this assumption on state-of-the-art papers~\cite{mcmahan2016communication, blanchard2017machine, mhamdi2018hidden} Although IID assumption leads to a better setting to study the impact of model utility requirement~\ref{req:p_utility}, in real world federated learning applications, $D_{c_i}$'s are typically unbalanced and non-IID. Therefore, we evaluated the performance of \ourmethod and \ourpattern in non-IID scenarios, and concluded that we met requirements~\ref{req:w_ownership} and~\ref{req:p_utility} with a slight increase in communication and computational overhead. Detailed experimental results can be found in Appendix~\ref{ssec:WafflenonIID}.}

While we focused on a single owner $O$, \ourmethod can also be used if several \client s would all be owners of the trained model. In this case, \agg\ can generate and distribute different subsets of the watermark set to each client for individual demonstration of ownership. There are challenges to be addressed in the case of collective ownership since the size of the watermark set increases linearly with the number of owners and there might be a decrease in utility. We consider extending \ourmethod to this case as future work. We will also explore how malicious clients can try to recover and degrade watermarks \textit{during} the training phase of the federated learning and provide detection/mitigation techniques against these adversaries.
\section{Related Work} \label{related}

Watermarking DNNs for ownership verification was first proposed in~\cite{uchida2017embeeding} by using backdooring techniques. Deepmarks~\cite{chen2019deepmarks} presents a collusion-secure watermarking method that encodes watermarks into the probability density function of weights using a specific regularization loss during the training phase. However, these techniques require direct access to model weights (white-box access) for ownership verification. DeepSigns~\cite{rouhani2018deepsigns} is the first watermarking method applicable with both white-box and black-box access, and it embeds watermarks to the activation maps of selected layers. DeepIPR~\cite{fan2019rethinking} proposes a passport-based DNN ownership verification scheme 
that tries to embed watermarks into a special passport layer of DNNs. Although DeepIPR is robust against watermark removal attacks, it is expensive and imposes a significant computational cost. 
A zero-bit watermarking algorithm~\cite{merrer2017adversarial} embeds watermarks into models, that are stolen via black-box methods, by leveraging adversarial perturbations. Although this approach is feasible, it heavily depends on adversarial examples and their transferability property across different models. 
Authors in~\cite{li2019prove} propose a blind-watermark based approach that inserts a specifically designed logo into the original sample via encoder such that the resulting sample is almost indistinguishable from the original one. All these proposals require full control of the training process and cannot be applied in federated learning.

       
Federated learning is vulnerable to adversarial attacks that alters the training and inference phase of the system. Poisoning attacks are first introduced in~\cite{bagdasaryan2018backdoor}, where a malicious client trains its local model on the backdoor and attempts to replace the global model with the poisoned model. Another powerful attack is model update poisoning attacks, where the adversary aims to prevent the global model from converging to a desirable state by sending poisoned model updates to the server~\cite{mhamdi2018hidden}. ~\cite{chen2017distributed} states that clients might suffer Byzantine failures, which leads to arbitrary behavior across communication rounds affecting the convergence of the global model. 

Client-server federated learning also suffers from privacy leakage~\cite{hitaj2017deep,melis2019exploiting,nasr2018comprehensive}. For example, in~\cite{hitaj2017deep}, a malicious client can learn about class representatives of other clients' training data by using generative adversarial networks. Attackers in client-server federated learning implement passive~\cite{nasr2018comprehensive} and active~\cite{melis2019exploiting} membership inference attacks in order to detect whether a sample belongs to the overall training set or a specific client. In passive attacks, the attacker could be the aggregator that can only observe individual model updates. In active attacks, malicious clients try to influence the global model in order to extract more information about other clients' training dataset. There have been a few attempts to prevent the privacy leakage applying differential privacy~\cite{geyer2017differentially} or using trusted execution environments in clients' devices~\cite{mo2020darknetz}. However, these methods trade-off either computational overhead or performance for privacy.

\bibliographystyle{IEEEtran}
\bibliography{reference}
\appendix

\section{Appendix}\label{sec:appendixa}
\renewcommand{\thefigure}{A\arabic{figure}}
\setcounter{figure}{0}
\renewcommand{\thetable}{A\arabic{table}}
\setcounter{table}{0}

\subsection{Algorithm for \ourmethod}\label{ssec:algorithmWaffle}

\begin{algorithm}[ht]
\caption{Functions and notations used in \ourmethod}
\resizebox{\columnwidth}{!}{
\begin{tabular}{ll}
  $t$ & aggregation round \\
  $w_{G(0)}$ &randomly initialized global model \\
   $w^-_{G(t+1)}$ & updated global model at $t$\\
   $w^+_{G(t+1)}$& watermarked global model at $t$\\
   $\mathrm{WM}_{w_G}$& watermark set designed by $O$\\
  $th$& min threshold value for $Acc(w^-_{G(t+1)}, \mathrm{WM}_{w_G})$\\
  $E_r$& max number of re-training rounds\\
  $E_i$& number of pre-training rounds\\
  $b_G$& mini batch of \agg \\
  $\eta_G$& learning rate in \textsc{Retrain}\\
  $\eta_i$& learning rate in \textsc{Pretrain}\\
  $\ell_G$& loss function\\
  $\triangledown \ell_G$& gradient of the loss function $\ell_G$\\
\end{tabular}}
\label{alg:algorithmPR}
\begin{algorithmic}[1]
\hrule
\STATE{{\textsc{Pretrain}$( w_{G(0)}, \mathrm{WM}_{w_G} ):$}}
    \FOR{round $i =1\cdots E_i$}{
     \FOR{$b_G\in \mathrm{WM}_{w_G}$}
     \STATE{ $w^+_{G(0)} \gets w_{G(0)} - \eta_i \triangledown \ell_G(w_{G(0)}; b_G)$}
      \ENDFOR}
      \ENDFOR
      \RETURN $w^+_{G(0)}$
 \end{algorithmic}
\hrule
\begin{algorithmic}[1]
 \STATE{\textsc{Retrain}$(w_{c_i(t)},\forall c_i \in C_{sub}):$}
    \STATE{$w^{-}_{G(t+1)}\gets FedAvg(w_{c_i(t)})$}
     \STATE $tr \gets 0$
     \WHILE{$Acc(w^-_{G(t+1)}, \mathrm{WM}_{w_{G}}) < th$ and $t_r \leq E_r$}{
      \FOR{$b_G\in \mathrm{WM}_{w_{G}}$}
      \STATE{$w^+_{G(t+1)} \gets w^-_{G(t+1)} - \eta_G \triangledown \ell_G(w^-_{G(t+1)}; b_G)$}
       \ENDFOR}
        \STATE $t_r \gets t_r + 1$
        \ENDWHILE
        \RETURN $w^+_{G(t+1)}$
 \end{algorithmic}
\end{algorithm}
 
\subsection{Detailed Experimental Setup and Hyperparameter Selection}\label{ssec:design}

To facilitate the comparative performance evaluation of \ourmethod\ (and its variants) and state-of-the-art watermark generation techniques, we used the following experimental setup: PyTorch (version 1.4.0)~\cite{paszke2017automatic} and Pysyft (0.1.21a1)~\cite{ryffel2018generic} library. Pysyft is a secure and private machine learning library that is used for applications including federated learning. All experiments are done in a computer with 2x12 core Intel(R) Xeon(R) CPUs (32GB RAM) and NVIDIA Quadro P5000 with 16GB memory. 

For all models and experiments, total number clients involved in the training is 100. In each aggregation round $t$, \aggregator\ averages model updates uploaded from 10 randomly selected $c_i$ to recompute a new global model $w_{G(t+1)}$. For all experiments, we used the cross-entropy loss, Stochastic Gradient Descent (SGD)~\cite{lecun1998gradient}
with a batch size of $50$ and learning rate of $0.1$ and $0.01$ for training MNIST and CIFAR10 models, respectively. We used a learning rate of $0.001$ at only one experiment to protect the model from diverging: watermarking MNIST models using unStruct.

During \textsc{Pretrain} and \textsc{Retrain}, we trained models using the cross entropy loss and SGD with a batch size of $50$. During \textsc{Pretrain}, we used a learning rate of $0.1$ and $5\times10^{-4}$ for MNIST and CIFAR10, respectively. During \textsc{Retrain}, we used a small learning rate of $0.005$ and $5\times10^{-4}$ for MNIST and CIFAR10 to ensure that the shift in the local minimum is as small as possible so that optimizing the model on the watermark set does not corrupt the actual task. We also used a momentum of $0.5$ and a weight decay of $5\times10^{-5}$ in \textsc{Pretrain}. During \textsc{Pretrain}, we trained MNIST models for 25 epochs with \ourpattern, 90 epochs with Embedded Content, 80 epochs with unRelate and 150 epochs with unConstruct. Similarly, we pre-trained CIFAR10 models for 30 epochs with \ourpattern, 60 epochs with Embedded Content, 55 epochs with unRelate and 200 epochs with unConstruct. For all experiments, the stop condition for \textsc{Retrain} is $Acc(w^-_{G(t+1)}, \mathrm{WM}_{w_G}) < 98\%$ or the maximum number of training rounds $E_r=100$ is reached.

\subsection{Generating State-of-the-art Watermark Sets}\label{app:generatewmsets}
\textbf{Embedded Content}~\cite{zhang2018protecting} takes a subset of the training data and modifies samples of this subset by adding a meaningful content (e.g., logo, text, a specifically designed pattern) into them. While original samples are labeled correctly, modified samples have incorrect labels that are pre-defined by the \modelowner. We construct Embedded Content by randomly selecting 10 images from the training data for each class, 100 images in total. We modify these images by adding a specifically designed pattern to them and assign incorrect labels to this modified set. We emphasize that Embedded Content requires training data knowledge, so it is not a suitable method for watermarking large-scale federated learning models. Nevertheless, we compare \ourpattern\ to Embedded Content, since it is proved to be a robust watermark~\cite{zhang2018protecting} against post-processing techniques. \textbf{unRelate}~\cite{zhang2018protecting,adi2018turning} is either designed as unstructured abstract images~\cite{adi2018turning} or as structured images from another data distribution~\cite{zhang2018protecting}, both of which are unrelated to the original task. These images are labeled with classes from the original task. For example, if the task is face recognition, the \modelowner\ might use different handwriting images to construct the watermark. We construct unRelate by randomly sampling 100 images from the ImageNet dataset which are unrelated to CIFAR10 and MNIST~\cite{zhang2018protecting}. \textbf{unStruct}~\cite{rouhani2018deepsigns} generates a random watermark set and adds it into different layers of DNN models. In order to imitate unStruct, we produce watermark samples with purely Gaussian noise as in~\cite{namba2019robust}. In both unRelate and unStruct, 10 classes are assigned to randomly selected 10 images, 100 in total.

\subsection{Additional Experimental Results}\label{app:additionalexperimentalresults}

\subsubsection{Fine-tuning attack}\label{ssec:finetuningappendix}

\begin{figure}[t]
    \centering
        \subfloat
        {\includegraphics[width=0.90\columnwidth]{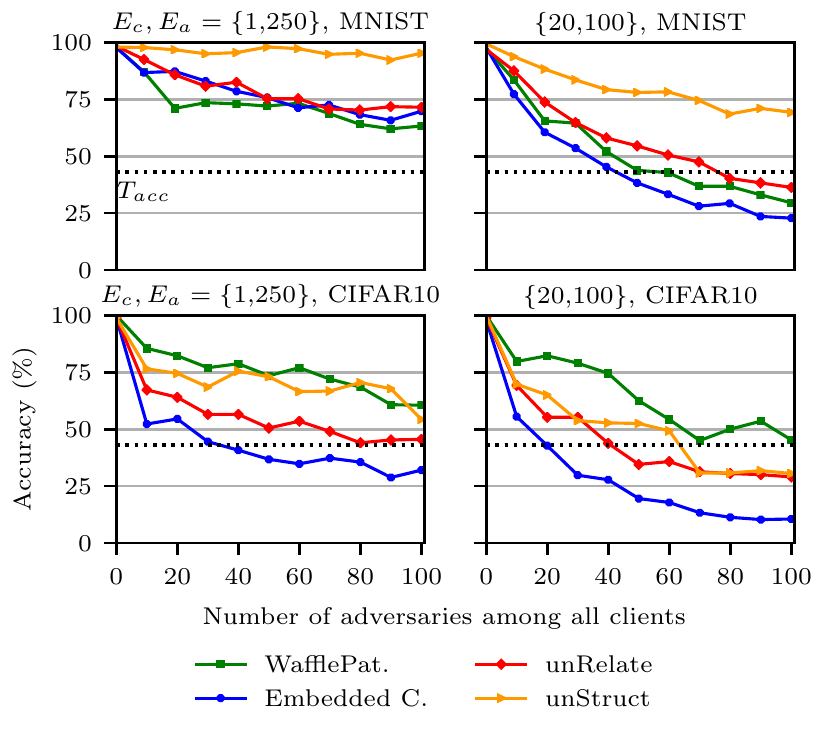}}
       \caption{Comparison of watermark accuracy for MNIST and CIFAR10 at different $\{E_c, E_a\}$ tuples when fine-tuning is implemented by an increasing number of malicious clients.}\label{fig:finetuningresults2}
\end{figure}

\begin{table}[t]
\caption{Test and watermark (WM) accuracy (\%) for different numbers of fine-tuning epochs on watermarked MNIST and CIFAR10 models using \ourpattern\ at different $\{E_c, E_a\}$ tuples. 1 adversary out of 100 clients.}\label{tb:finetuningtrendMNISTCIFAR2}
\centering
\resizebox{\columnwidth}{!}{
\begin{tabular}{c|c|c|c|c|c|c|c|c}
\hline
& \multicolumn{4}{|c|}{MNIST} & \multicolumn{4}{|c}{CIFAR10} \\  \hline
\multicolumn{1}{c|}{\footnotesize $\{E_c, E_a\}$} &  \multicolumn{2}{c|}{ $\{1,250\}$} &  \multicolumn{2}{c|}{$\{20,100\}$} &  \multicolumn{2}{c|}{$\{1,250\}$} &  \multicolumn{2}{c}{$\{20,100\}$} \\
 \hline
epoch &\hfil Test & \hfil WM& \hfil Test & \hfil WM & \hfil Test & \hfil WM & \hfil Test & \hfil WM \\ \hline
 0    & \hfil 98.8& \hfil 99.0& \hfil 98.8  & \hfil 99.0 & \hfil  85.7 &\hfil 100.0 & \hfil  85.6  & \hfil 100.0\\
 
20   & \hfil 98.6& \hfil 97.8& \hfil 98.5  & \hfil 88.2 & \hfil  85.4 & \hfil 94.2 & \hfil  84.4 & \hfil 95.2 \\

40   & \hfil 98.6& \hfil 97.2& \hfil 98.5  & \hfil 88.0 & \hfil  85.4 & \hfil  94.0 & \hfil 84.5 & \hfil 94.8 \\

60   & \hfil 98.6& \hfil 95.5& \hfil 98.5  & \hfil 88.0 & \hfil  85.4 & \hfil 93.0 & \hfil 84.6 & \hfil 95.0 \\

80   & \hfil 98.6& \hfil 95.2& \hfil 98.5  & \hfil 88.0 & \hfil  85.4 & \hfil 93.0 & \hfil  84.6 & \hfil 95.0 \\

100 & \hfil 98.6& \hfil 95.0& \hfil 98.5  & \hfil 88.0 & \hfil  85.4 & \hfil  93.0 & \hfil  84.6 & \hfil 94.8 \\ \hline
 \end{tabular}}
\end{table}

{Fine-tuning is the most likely attack that an adversary might attempt, since it only involves re-training the model on its original training dataset. We tested fine-tuning by making adversaries run the same local training procedure as used during federated learning once more on the final watermarked model $w_{adv}={w^+_{G(t)}}$.
\newtext{Figure~\ref{fig:finetuningresults2} shows that all watermark sets except Embedded Content are robust to fine-tuning even when up to 50\% of clients are malicious.} Table~\ref{tb:finetuningtrendMNISTCIFAR2} further shows the evolution of the test and watermark accuracy when running an increasing number of fine-tuning epochs against \ourpattern. We can see that the watermark accuracy initially decreases by at most 10 percentage points and then reaches a plateau after a maximum of 100 epochs. The final watermark accuracy is always high enough (\textgreater 85\%) to enable reliable proof of ownership showing that \ourpattern is resilient to a large number of fine-tuning epochs.

\subsubsection{Fine-pruning attack}\label{ssec:finepruningappendix}
\begin{figure}[t]
    \centering
         \subfloat
         {\includegraphics[width=0.90\columnwidth]{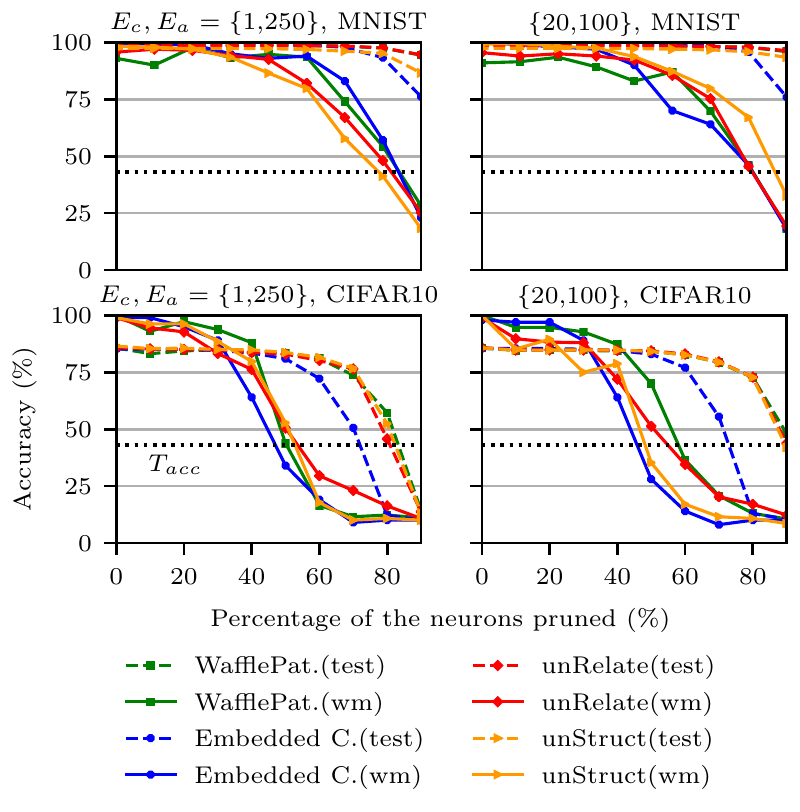}}
       \caption{Comparison of the test and watermark (wm) accuracy for MNIST and CIFAR10 at different $\{E_c, E_a\}$ tuples when the pruning attack is implemented by one adversary.}\label{fig:pruningresultsPR2}  
\end{figure}
 \begin{figure}[t]
    \centering
        \subfloat[MNIST\label{fig:mnistncleansePR}]{\includegraphics[width=0.90\columnwidth]{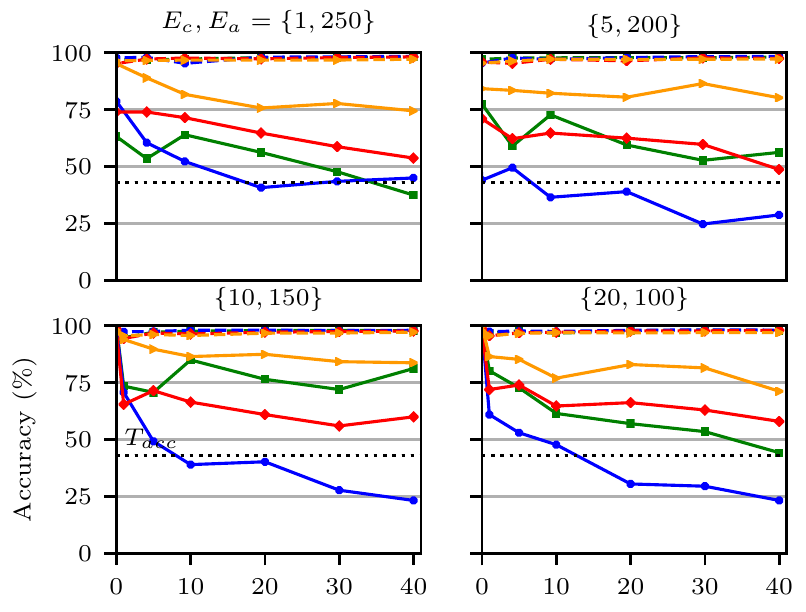}}
        \qquad
         \subfloat[CIFAR10\label{fig:cifarncleansePR}]{\includegraphics[width=0.90\columnwidth]{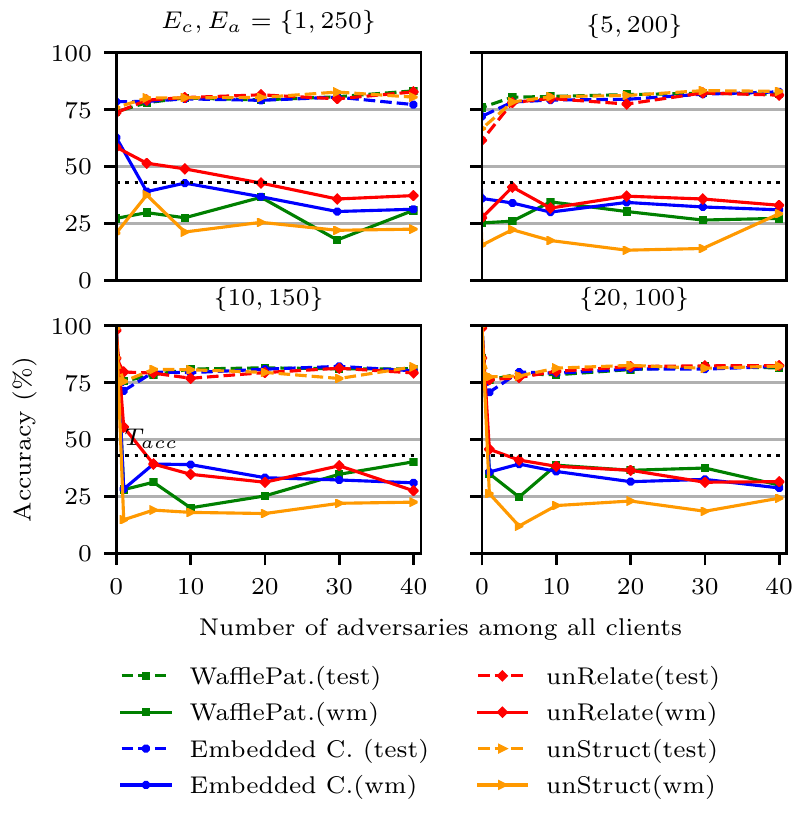}}
       \caption{Comparison of the test and watermark (wm) accuracies for the MNIST (a) and CIFAR10 (b) models when patching via unlearning is implemented against watermarked models.}\label{fig:ncleanseresultsPR}   
\end{figure}
The pruning technique in~\cite{han2015learning} first removes a number of connections with magnitudes close to zero from the dense model and obtains a more sparse model. Then, it re-trains this sparse model to recover the test accuracy. In our setup, adversaries prune their model $w_{adv}={w^+_{G(t)}}$ and then implement the fine-tuning attack. Figure~\ref{fig:pruningresultsPR2} illustrates the test and watermark accuracy of $w_{adv}$ when pruning is implemented by one adversary using different pruning rates. \newtext{We see that the pruning is effective at decreasing the watermark accuracy for all watermark sets when a high percentage of neurons are removed from the model (70\% for MNIST and 50\% for CIFAR10). However, as explained in Section~\ref{sssection:pruning}, when the watermark accuracy is below $T_{acc}$, the test accuracy also starts to decrease and the pruned model cannot achieve a high test accuracy as the original model $w_{adv}$.}

\subsubsection{Neural Cleanse attack}\label{app:neuralcleanse}
Table~\ref{tb:anomalyindex} shows that in MNIST, Neural Cleanse returns an anomaly index below 2.0 for all watermarked models using all four watermark sets. In CIFAR10, we measure an anomaly index around 2.5, but Figure~\ref{fig:reversedtriggers} shows that the reversed triggers of marked classes and original watermarks are dissimilar. Patching-via-unlearning results are plotted in Figure~\ref{fig:ncleanseresultsPR}. As can be seen from figure, in MNIST, \ourpattern is robust when less than 40\% of clients are adversaries. In CIFAR10, one adversary might evade the verification but the performance drop is more than 5\%.

 \begin{figure}[t]
    \centering
     \subfloat[Clean \label{fig:orig1mnist}]{\includegraphics[width=0.20\linewidth]{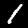}}
       \quad
       \subfloat[WM\label{fig:wm1mnist}]{\includegraphics[width=0.20\linewidth]{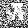}}
       \;
        \subfloat[Reversed\label{fig:nc1mnist}]{\includegraphics[width=0.20\linewidth]{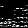}}
        \\
        \subfloat[Clean \label{fig:orig2cifar}]{\includegraphics[width=0.20\linewidth]{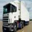}}
        \;
        \subfloat[WM\label{fig:wm2cifar}]{\includegraphics[width=0.20\linewidth]{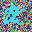}}
        \;
        \subfloat[Reversed\label{fig:nc2cifar}]{\includegraphics[width=0.20\linewidth]{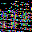}}
       \caption{Visualization of reversed triggers where Neural Cleanse is implemented by one adversary against watermarked models using \ourpattern. We show the original images (a) MNIST class 1 and (d) CIFAR10 class 9; example watermark samples for them (b) and (e) respectively; and reversed triggers via Neural Cleanse (c) and (f) respectively.} \label{fig:reversedtriggers}   
\end{figure}

\subsubsection{Communication overhead}\label{app:commoverhead}

Figure~\ref{fig:commoverhead2} shows the convergence in test accuracy for MNIST and CIFAR10 during the training at different $\{E_c, E_a\}$ tuples.  \newtext{In MNIST, unStruct has the slowest convergence rate, while Embedded Content converges slightly slower than other watermarks in CIFAR10 except $\{E_c, E_a\} = \{5,200 \}$. Embedded Content has also the second slowest convergence rate in MNIST.} \ourpattern satisfies \ref{req:p_communication} in both MNIST and CIFAR10. 

%

\begin{table}[t]
\caption{ Average anomaly index for watermarked models watermarked using different watermark sets. Results are averaged over watermarked models with different $\{E_c, E_a\}$ tuples and various number of adversaries.}\label{tb:anomalyindex}
\centering
\resizebox{0.95\columnwidth}{!}{
\begin{tabular}{ p{\dimexpr 0.18\linewidth-2\tabcolsep}
                       | p{\dimexpr 0.20\linewidth-2\tabcolsep}
                       | p{\dimexpr 0.20\linewidth-2\tabcolsep}
                       | p{\dimexpr 0.20\linewidth-2\tabcolsep}
                       | p{\dimexpr 0.20\linewidth-2\tabcolsep}}
\hline
\multicolumn{1}{c|}{Dataset} &  \multicolumn{1}{|c|}{\textbf{\ourpattern}} & \multicolumn{1}{|c|}{\textbf{Embedded C.}} & \multicolumn{1}{|c|}{\textbf{unRelate}} & \multicolumn{1}{|c}{\textbf{unStruct}} \\ \hline 
\textbf{MNIST} & \hfil 1.27 & \hfil 1.32 & \hfil 1.47 & \hfil 1.54 \\ 
\textbf{CIFAR10} & \hfil 2.35 & \hfil 2.16 & \hfil 2.32 & \hfil 2.15 \\ \hline
\end{tabular}}
\end{table}

\begin{figure}[t]
     \centering
        \subfloat{\includegraphics[width=0.80\columnwidth]{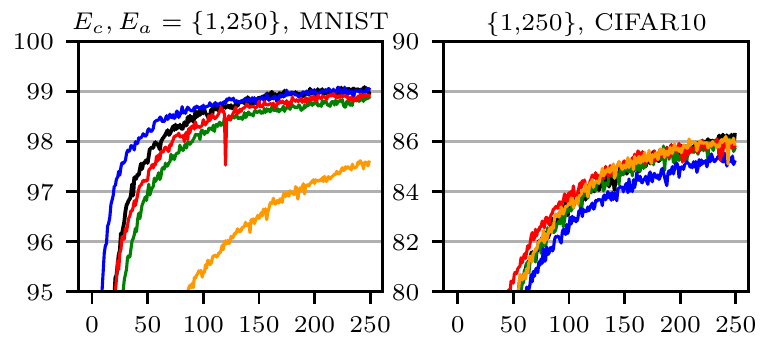}}
        \quad
        \subfloat{\includegraphics[width=0.80\columnwidth]{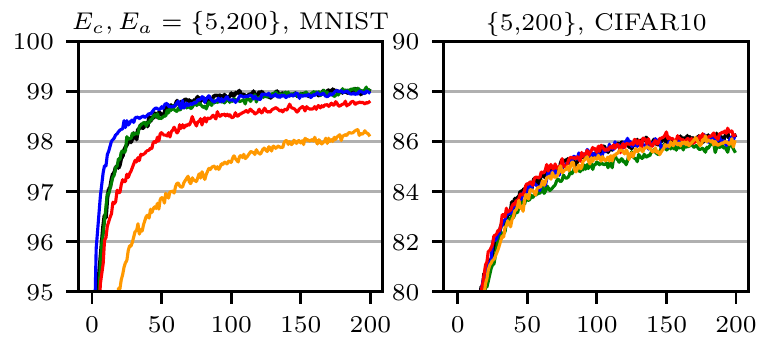}}
        \quad
         \subfloat{\includegraphics[width=0.80\columnwidth]{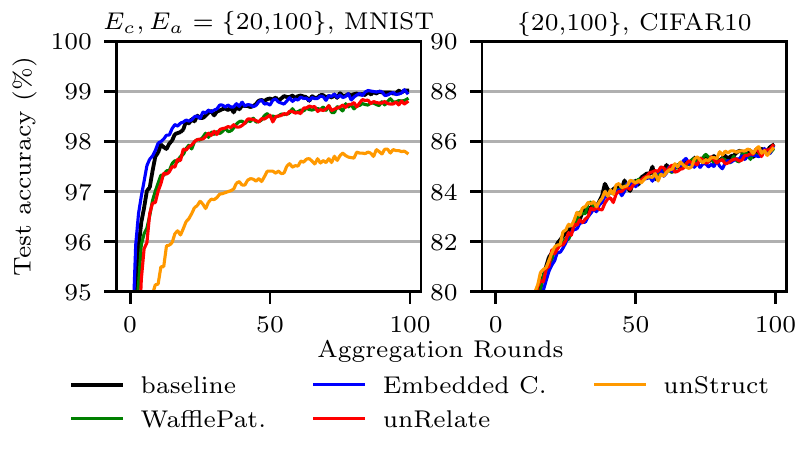}}
    \caption{Progression of the test accuracy for MNIST and CIFAR10. Baseline and watermarked models using different watermark sets are shown in the figure.}\label{fig:commoverhead2} 
\end{figure}

\subsection{\ourmethod in Non-IID scenarios}\label{ssec:WafflenonIID}
\newtext{As explained in Section~\ref{ssec:discussion}, we evaluated the performance of \ourmethod and \ourpattern in a more realistic federated learning setup, where each client holds a non-IID training data. For both MNIST and CIFAR10, we implemented the same partitioning in~\cite{mcmahan2016communication} in order to distribute the training set to clients in a non-IID fashion. In this partitioning, clients and adversaries will have samples for only two classes, so each resulting $D_{c_i}$ is still balanced (i.e. includes the same number of samples for each class) but highly non-IID.}

\newtext{We trained both baseline and watermarked models using \ourpattern, where the learning rate is decreased by 2 and $E_a$ is increased in order to obtain a model with a reasonable performance and prevent the loss from diverging. As shown in Table~\ref{tb:noniid}, the test accuracy of baseline and watermarked models are quite similar while the watermark accuracy is still high. Therefore, \ourmethod with \ourpattern satisfies demonstration of ownership~\ref{req:w_ownership} and model utility~\ref{req:p_utility} requirements in extreme non-IIDness. However, the computational overhead is higher in CIFAR10, since the convergence to both training and watermark sets slows down due the skewness in client's data distribution and the watermark set. We also should note that the extreme non-IIDness also negatively affects the watermark removal techniques presented in Section~\ref{ssec:wmrobustness}. For example, in Neural Cleanse, one $adv$ can only find reverse triggers for only two classes, which reduces the effect of patching-via-unlearning. In pruning, one $adv$ retrains the pruned model with only samples from two classes, and thus cannot recover the test accuracy as much as in the IID scenario.}

\begin{table}[t]
\caption{\newtext{Test accuracy (\%), watermark accuracy (\%) and computational overhead (\%) of watermarked models using \ourpattern.}}\label{tb:noniid}
\centering
\begin{tabular} {c|c|cccc}
\hline
& Baseline models & \multicolumn{3}{c}{Watermarked models} \\ \hline
\bf{MNIST}& Test & Test &WM &  Comp.  \\ 
$\{E_c, E_a \}$ & Acc. & Acc. & Acc. & overhead  \\ \hline
$\{1, 500\}$ & 98.44 & 97.61 & 100.0 & 0.80\\
$\{1, 450\}$ & 98.49 & 98.60 & 100.0 & 0.23\\
$\{1, 300\}$ & 98.14 & 98.38 & 100.0 & 0.16\\
$\{1, 350\}$ & 98.51 & 98.74 & 100.0 & 0.15\\ \hline

\bf{CIFAR10}& Test & Test &WM & Avg comp.   \\ 
$\{E_c, E_a \}$ & Acc. & Acc. & Acc. & overhead  \\ \hline
$\{1, 450\}$ & 81.88 &  81.97 &  99.00 & 3.24 \\
$\{1, 400\}$ & 83.30 &  83.24 & 100.0  & 4.17\\
$\{1, 350\}$ & 83.49 &  82.76 &  99.00 & 4.02\\
$\{1, 300\}$ & 81.28 &  80.59 &  100.0 & 2.04\\ \hline
\end{tabular}
\end{table}}

\end{document}